\documentclass[aps,pra,10pt,twocolumn,superscriptaddress,floatfix]{revtex4-2}
\usepackage{chngcntr}
\usepackage{epstopdf}
\usepackage{graphicx}
\usepackage{dcolumn}
\usepackage{bm}
\usepackage{bbm,bbold}
\usepackage{color}
\usepackage{xcolor, soul}
\sethlcolor{green}
\usepackage{amsfonts}
\usepackage{amsmath}
\usepackage{mdframed}
\usepackage[normalem]{ulem}
\usepackage{mathrsfs}
\usepackage[percent]{overpic}

\usepackage{subfigure}
\usepackage{braket}
\usepackage[colorlinks=true,citecolor=blue]{hyperref}
\hypersetup{colorlinks=true,citecolor=blue,linkcolor=blue,urlcolor=blue}

\begin{document}
	\title{Entanglement and steering in quantum batteries}
	\author{Dayang Zhang}
	\affiliation{Department of Physics, Zhejiang Sci-Tech University, Hangzhou 310018, China}
	\author{Shuangquan Ma}
	\affiliation{Department of Physics, Zhejiang Sci-Tech University, Hangzhou 310018, China}
	\author{Yunxiu Jiang}
	\affiliation{Department of Physics, Zhejiang Sci-Tech University, Hangzhou 310018, China}
	\author{Youbin Yu}
	\email{ybyu@zstu.edu.cn}
	\affiliation{Department of Physics, Zhejiang Sci-Tech University, Hangzhou 310018, China}
	\author{Guangri Jin}
	\affiliation{Department of Physics, Zhejiang Sci-Tech University, Hangzhou 310018, China}
	\author{Aixi Chen}
	\affiliation{Department of Physics, Zhejiang Sci-Tech University, Hangzhou 310018, China}
	\begin{abstract}
		The advantage of quantum batteries is that quantum resources can be used to improve charging efficiency. The quantum resources that are known to be available are: quantum entanglement and quantum coherence. In this paper, we introduce quantum steering as a new quantum resource into batteries for the first time. We analyze the relationship between quantum steering, quantum entanglement, energy storage, and extractable work by considering two models: Field-quantum battery and Cavity-Heisenberg quantum battery. We find that in the steerable range, the quantum steering of different qubits has a maximum or minimum value, which corresponds to the energy storage of the battery, and the extractable work has a maximum value. The occurrence of the minimum value of quantum entanglement is always accompanied by the occurrence of the maximum value of parameters such as energy storage. Ultimately, we analyzed the reasons for these results using the purity of the system. And found a relatively general conclusion: when the purity is at the maximum, important parameters such as the energy storage of the battery are also at the maximum.
		
	\end{abstract}
	
	\maketitle
	\section{Introduction}
	In recent years, with the rapid development of quantum informatics and quantum thermodynamics, many interdisciplinary products have emerged. These include quantum heat engines\cite{pirandola2018advances,PhysRevLett.123.230502,PhysRevE.76.031105,PhysRevE.79.041129,PhysRevLett.123.240601,PhysRevResearch.2.032062}, quantum sensors\cite{degen2017quantum,pirandola2018advances,marciniak2022optimal,PhysRevLett.123.230502,boss2017quantum,bass2024quantum}, quantum batteries(QBs)\cite{PhysRevLett.122.210601,PhysRevLett.118.150601,jiang2022quantum,alicki2013entanglement,carrasco2110collective,dou2022cavity,zhang2019powerful,PhysRevLett.131.030402,pirmoradian2019aging,carrega2020dissipative,qu2023experimental,joshi2022experimental,le2018spin}, and more. As an outstanding product with quantum effect, QBs have attracted widespread attention. Unlike batteries in the traditional sense, it is considered to be a device for energy storage with two-level atoms\cite{ferraro2018high}. It has the advantages of fast charging speed and high stability. Up to now, people have paid more attention to its charging process, including the utilization of energy storage\cite{PhysRevE.102.052109,PhysRevB.104.245418,andolina2019extractable,shi2022entanglement}, stable charging\cite{PhysRevLett.122.210601,PhysRevE.104.044116,PhysRevA.109.022226}, fast charging\cite{ferraro2018high,PhysRevB.102.245407,dou2022cavity,carrasco2110collective}, etc. The concept of QBs was first developed by Alicki and Fannes proposed that they show that QBs can use quantum entanglement to improve extractable work compared to ordinary batteries\cite{alicki2013entanglement}. After that, Ferraro et al. used collective charging to prove that QBs charge faster than ordinary batteries, and the charging rate is $N^{1.5}$ times that of the charging cell, and the more charging cells, the faster the charging rate\cite{ferraro2018high}. Crescente et al. adopted the two-photon charging mode to break through the charging rate of QBs to $N^{2}$ times\cite{PhysRevB.102.245407}. Immediately afterwards, Zhang propose an efficient charging quantum
	battery achieved by considering an extension Dicke model\cite{dicke1954coherence,baumann2010dicke} with dipole-dipole interaction and an external driving
	field. It shows that by optimizing the parameters, the efficiency of the battery can rise to $N^{1.6}$ times\cite{PhysRevE.107.054125}. Dou et al. put the Heisenberg chain into a cavity for charging\cite{dou2022cavity}, which also demonstrated that collective charging has and suggested that this advantage may be related to quantum entanglement\cite{RevModPhys.81.865} and quantum coherence\cite{RevModPhys.89.041003}. Carrasco takes into account QBs that charge in a dissipative environment, and the results show that collective charging in a dissipative environment is also advantageous\cite{carrasco2110collective}. 
	
	For the extraction of work in batteries, Andolina made a study. The results show that as the charging cell $N\rightarrow \infty$, the work in the battery can be completely extracted, so that the conversion efficiency is $1$\cite{PhysRevLett.122.047702}. Garcia Pintos' work proved that quantum coherence can enhance the charging process \cite{PhysRevLett.125.040601}. Shi et al. likewise demonstrated this. At the same time, shi also shows that the conversion efficiency of the extractable work in the pure state of the system is $1$.\cite{shi2022entanglement}. Liu, using a central spin QB, proved that entanglement is negatively correlated with extractable work\cite{Liu_2020}. Kamin et al. showed that entanglement cannot be used to improve charging efficiency, and coherence in the system is an important resource to improve the charging process\cite{PhysRevE.102.052109}.
	
	All of the above studies focus on the role of quantum resources in accelerating the charging process of batteries. In addition to quantum entanglement and quantum coherence, we also know that quantum steering\cite{RevModPhys.92.015001,PhysRevLett.114.060403} is an important quantum resource. It is often used in quantum decryption\cite{PhysRevA.106.012433,PhysRevLett.123.170402,sun2017exploration}, quantum communication\cite{PhysRevA.88.062338,nagy2016epr,chiu2016no}, and other fields.
	
	However, is it also a resource that can be exploited in QB syetems?
	
	In order to introduce quantum steering into the battery system. In this paper, we use the principle of entropy uncertainty to discriminate and measure quantum steering\cite{PhysRevA.87.062103}. We use Field-QB and Cavity-Heisenberg chain QB to explore the effects of quantum entanglement and quantum guidance on important parameters such as battery energy storage. Our results show that the maximum value of quantum entanglement is negatively correlated for battery energy storage, etc. In the steerable range, the quantum steering maxima in different directions are positively or negatively correlated with battery energy storage, etc. Finally, we analyzed the system purity and found that the system purity is a key factor in causing this cause. If the purity is at the maximum, the energy storage effect of the battery will be better. In the pure state, the battery can reach a saturated state.
	
	\section{Charging protocols, energy storage, extractable work and resources in the charging process}
	
	We consider two quantum systems, $a$ and $b$. where $a$ is the “charger,” which can be either a cavity or a drive field. where $b$ is the proper “quantum battery(QB),” initially prepared in the ground state. The state of the whole system is represented by $\rho (t)$, and the Hamiltonian is represented by $H=H_a$+$H_b$. Energy storage of a battery is defined as\cite{PhysRevB.99.035421}:
	\begin{align}
		E(t)=\mbox{Tr}[H_{b}\rho(t)] -  \mbox{Tr}[H_{b} \rho(0)]
		\label{11}
	\end{align}
	where $\rho (t) =U(t)\rho(0)U(t)^{\dagger}$ with $U(t)=e^{-i\int_{0}^{t} {H}dt}$. Tr$[H_{b}\rho(t)]$ denotes the mean energy of the battery state $\rho_b(t)$ and $\rho_b(t)$=Tr$_a$[$\rho(t)]$\cite{PhysRevA.101.032115}. 
	
	We know that not all stored energy in a QB can be extracted, which is known as the second law of thermodynamics. A proper measure of the extractable work for
	the state $\rho_b(t)$ is provided by the ergotropy
	\begin{align}
		\varepsilon(t)=\mbox{Tr}[H_{b}\rho_b(t)]-\mbox{Tr}[H_{b}\overline{\rho_b(t)}]
		\label{7}
	\end{align}
	where $\overline{\rho_b(t)}$=$\sum_{n}r_{n}|\pi_{n}\rangle\langle\pi_{n}|$, the eigenstates of ${H}_b=\sum_n\pi_n|\pi_n\rangle\langle\pi_n|$ and $\rho_b(t)$=$\sum_nr_n|r_n\rangle\langle r_n|$ are reordered so that $r_{0}\geq r_{1}\geq r_{2}\geq\cdots $ and $ \pi_{0}\leq\pi_{1}\leq\pi_{2}\leq\cdots.$ This kind of quantum work has been
	experimentally measured recently in a single-atom heat
	engine\cite{van2020single} and a spin heat engine\cite{PhysRevLett.123.080602}.
	
	Efficiency during energy transfer is an important parameter in batteries. We are defined by the ratio of ergotropy to the energy stored in the battery:
	\begin{align}
		\eta(t)=\frac{\varepsilon(t)}{E(t)}
	\end{align}
	
	We focus on the role of quantum entanglement and quantum steering in battery systems. We can take a partial trace of the entire system to represent battery-battery entanglement or battery-charger entanglement. The corresponding von Neumann entropy is given by\cite{PhysRevLett.78.2275}
	\begin{align}
		S(t)=-\mathrm{Tr}(\rho_b(t)\log_2\rho_b(t)).
		\label{6}
	\end{align}
	where the magnitude of the entanglement varies with the dimension $d$ of space, and the maximum value does not exceed log$d$.
	
	Quantum steering we can follow the definition in \cite{PhysRevA.87.062103}, one can say that Alice can steer Bob’s state if the results
	presented by Alice and observed by Bob have correlations that do not admit a local-hidden-
	state (LHS) model. For states admitting LHS models, one can derive the LHS constraint \cite{PhysRevLett.110.130407},
	\begin{align}
		H(R^b|R^a)\geq\sum_\lambda P(\lambda)H_Q(R^b|\lambda).
	\end{align}
	Here,$R^a$ and $R^b$ denote the discrete observables for systems a and b, respectively, $H(R^b|R^a)$ is the conditional entropy aftermeasurements on the subsystem a, and $H_Q(R^b|\lambda)$is the discrete Shannon entropy of the probability distribution $P_{Q}(R^{\tilde{B}}|\lambda).$ On the other hand,
	any pair of discrete observables $R$ and $S$ in the same $N$-dimensional Hilbert space, with the eigenbases $\left|R_{i}\right\rangle $ and $|S_{i}\rangle(i=1,\ldots,N),$ admit the entropy uncertainty relation
	\begin{align}
		H_Q(R)+H_Q(S)\geq\log(\Omega),
	\end{align}
	where $\Omega\equiv\operatorname*{min}_{i,j}(1/|\langle R_{i}|S_{j}\rangle|^{2}).$ Using the above two equations, one immediately arrives at the entropy-based steering inequality,
	\begin{align}
		H(R^b|R^a)+H(S^b|S^a)\geq\log(\Omega^b),
	\end{align}
	where $\Omega^{b}$ is the value $\Omega$ associated with the observables $R^b$ and $S^b$.
	
	For spin systems, by considering the complete set of pairwise complementary Pauli operators $X$, $Y$ and $Z$, there exists the steering inequality,
	\begin{align}\label{3}
		I^{ab}=H(\sigma_x^b|\sigma_x^a)+H(\sigma_y^b|\sigma_y^a)+H(\sigma_z^b|\sigma_z^a)\geq2.
	\end{align}
	If this inequality is violated, we say that Alice can steer Bob.
	
	In this paper, we consider the dimension of the system $d$=$4$, and the density matrix of the corresponding system is:
	\begin{align}\label{1}
		\rho=\begin{pmatrix}\rho_{11}&\rho_{12}&\rho_{13}&\rho_{14}\\\rho_{21}&\rho_{22}&\rho_{23}&\rho_{24}\\\rho_{31}&\rho_{32}&\rho_{33}&\rho_{34}\\\rho_{41}&\rho_{42}&\rho_{43}&\rho_{44}\end{pmatrix},
	\end{align}
	To facilitate our local measurements throughout the system, Eq.(\ref{1}) can be expressed as: 
	\begin{align}\label{2}
		\rho=&\frac{1}{4}\Bigg(\boldsymbol{I}\otimes\boldsymbol{I}+\sum_{i}n_{i}\boldsymbol{\sigma}_{i}\otimes\boldsymbol{I}+\sum_{i}s_{i}\boldsymbol{I}\otimes\boldsymbol{\sigma}_{i}\notag\\&+\sum_{i}c_{i}\boldsymbol{\sigma}_{i}\otimes\boldsymbol{\sigma}_{i}\Bigg),
	\end{align}
	where ${\sigma}_{i}$ is a Pauli operator, $i$=$1$,$2$,$3$. With $n_1=\rho_{13}+\rho_{24}+\rho_{31}+\rho_{42}$, $n_2=i(\rho_{13}+\rho_{24}-\rho_{31}-\rho_{42})$, $n_3=\rho_{11}+\rho_{22}-\rho_{33}-\rho_{44}$, $s_1=\rho_{12}+\rho_{21}-\rho_{34}-\rho_{43}$, $s_2=i(\rho_{12}+\rho_{21}-\rho_{34}-\rho_{43})$, $s_3=\rho_{11}-\rho_{22}+\rho_{32}-\rho_{41}$, $c_1=\rho_{14}+\rho_{23}-\rho_{32}-\rho_{41}$, $c_2=\rho_{23}+\rho_{32}-\rho_{14}-\rho_{41}$, and  $c_3=\rho_{11}-\rho_{22}-\rho_{33}+\rho_{44}$.
	
	\begin{figure*}[t]
		\centering
		\includegraphics[width=0.65\linewidth]{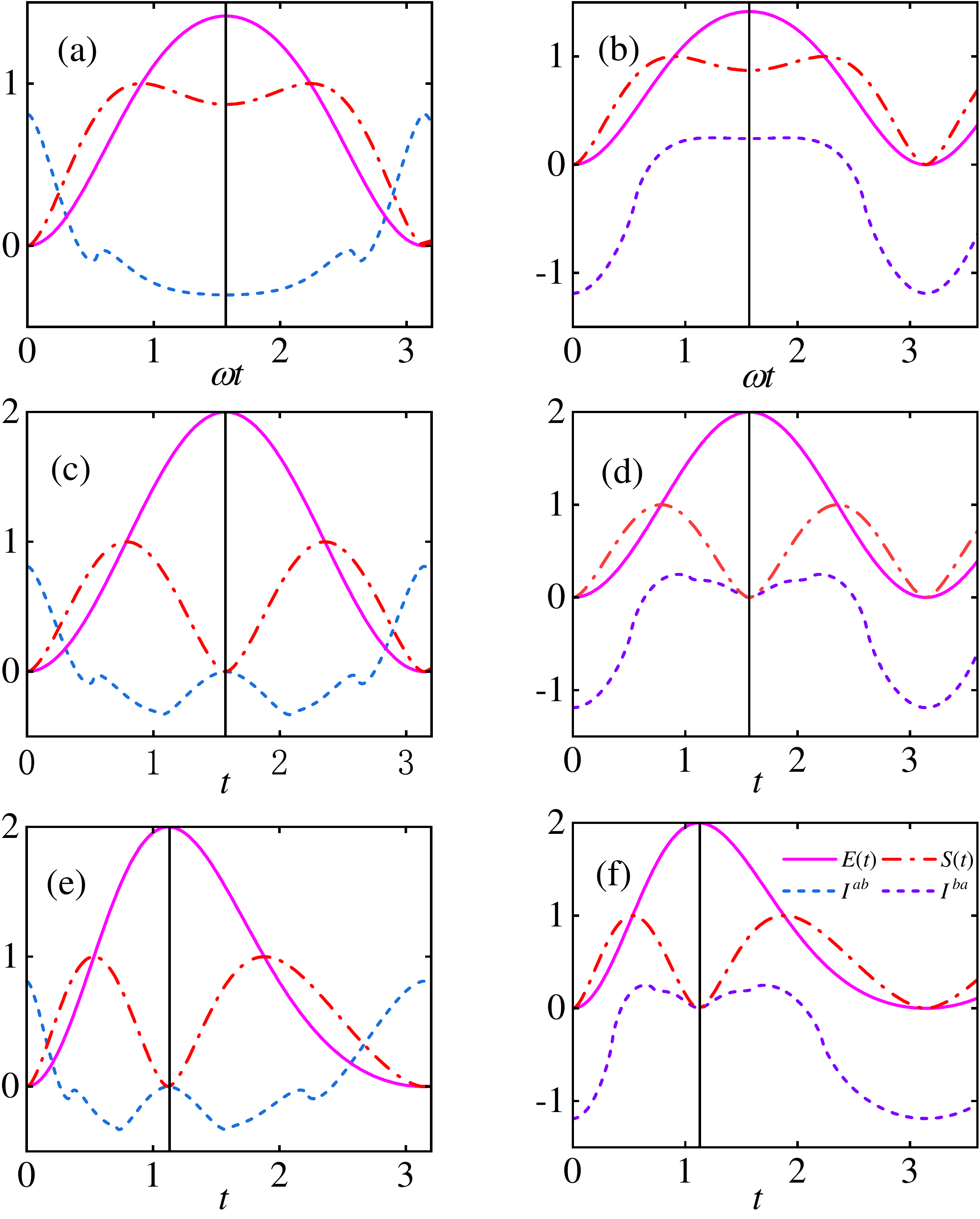}
		\caption{(a)-(f) plots the battery-battery entanglement $S(t)$, the $a\rightarrow b$ steering $I^{ab}$, $b\rightarrow a$ steering $I^{ba}$, and the stored energy $E(t)$ as a function of charging time $t$ for $\Delta=1$. (a) and (b) indicate that the harmonic field charges the battery, $A=2, B=0$. (c) and (d) indicate that the electrostatic field charges the battery, $A=0, B=2$. (e) and (f) indicate that the mixed field charges the battery, $A=1,B=2$.}
		\label{fig1}
	\end{figure*}
	
	For the spin state given by Eq.(\ref{2}), the quantity $I^{ab}$ of Eq.(\ref{3}) becomes
	\begin{align}
		I^{ab} =&\sum_{i=1,2,3}[(1+n_i)\log(1+n_i)+(1-n_i)\log(1-n_i) \notag\\&
		-\frac{1}{4}(1+c_i+n_i+s_i)\log(1+c_i+n_i+s_i)\notag\\&
		-\frac{1}{4}(1+c_i-n_i-s_i)\log(1+c_i-n_i-s_i) \notag\\&
		-\frac{1}{4}(1-c_i-n_i+s_i)\log(1-c_i-n_i+s_i) \notag\\&
		-\frac{1}{4}(1-c_i+n_i-s_i)\log(1-c_i+n_i-s_i)], 
		\label{9}
	\end{align}
	with the base of logarithms being $2$.
	
	For the steering from Bob to Alice, similar expressions read
	\begin{align}
		I^{ba} =&\sum_{i=1,2,3}[(1+s_i)\log(1+s_i)+(1-s_i)\log(1-s_i) \notag\\&
		-\frac{1}{4}(1+c_i+n_i+s_i)\log(1+c_i+n_i+s_i)\notag\\&
		-\frac{1}{4}(1+c_i-n_i-s_i)\log(1+c_i-n_i-s_i) \notag\\&
		-\frac{1}{4}(1-c_i-n_i+s_i)\log(1-c_i-n_i+s_i) \notag\\&
		-\frac{1}{4}(1-c_i+n_i-s_i)\log(1-c_i+n_i-s_i)], 
		\label{10}
	\end{align}
	
	\section{Field-QB}
	
	Now let's consider a QB that is charged by a hybrid field to the Heisenberg chain\cite{jiang2022quantum}. Hamiltonian can be written as
	\begin{align}
		H_a=(\frac{A\cos(\omega t)+B}{2})\sum_{i=1}^{2}\sigma _{i}^{x},
		\label{4}
	\end{align}
	\begin{align}
		H_{b} =&\frac{\Delta }{2}\sum_{i=1}^{2}\sigma _{i}^{z}+\frac{g}{4}%
		\sum_{i\neq j}^{2}(\sigma _{i}^{x}\sigma _{j}^{x}+\sigma _{i}^{y}\sigma
		_{j}^{y}+\sigma _{i}^{z}\sigma
		_{j}^{z})  \notag \\
	\end{align}
	\begin{figure*}[tb]
		\centering
		\includegraphics[width=0.85\linewidth]{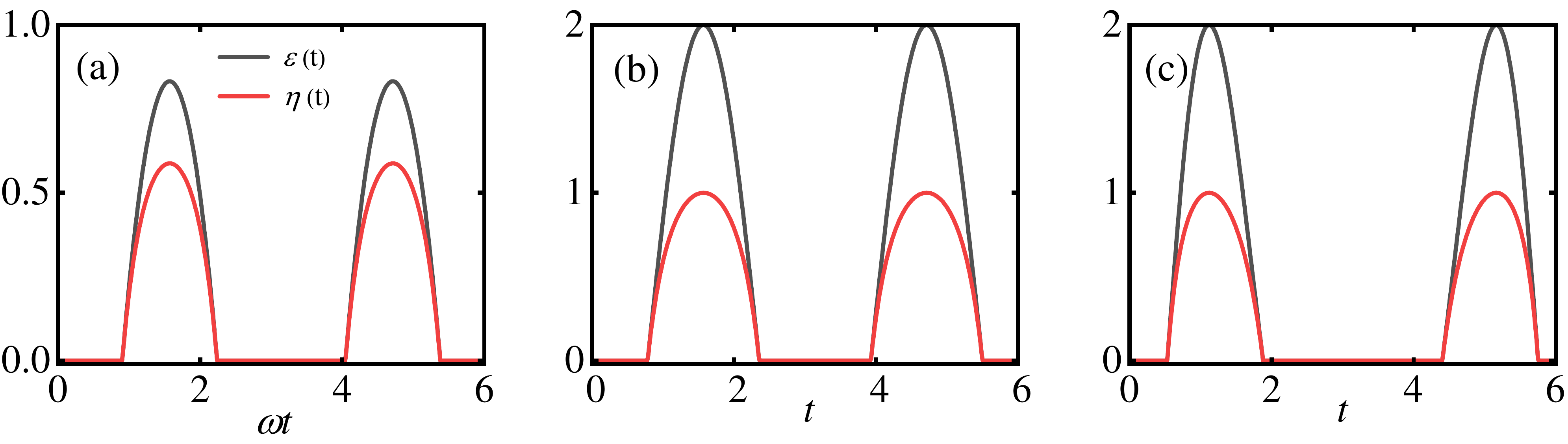}
		\caption{(a)-(c) plots the ergotropy $\varepsilon(t)$ and the efficiency of the conversion of work $\eta(t)$ as a function of charging time $t$ for $\Delta=1$. (a), (b), and (c) represent the situation in which a harmonic field($A=2,B=0$), an electrostatic field($A=0,B=2$), and a mixed field($A=1,B=2$) charge the battery, respectively.}
		\label{fig2}
	\end{figure*}
	where $A$ and $B$ are the driving amplitudes of the harmonic field and electrostatic field, respectively. $\omega$ is the modulated frequency.
	$g$ is the atom–atom coupling strength, including the repulsive ($g > 0$) and attractive ($g < 0$) interactions. 
	From Eq.(\ref{4}) we know that when $A=0$, it will be driven by a electrostatic field. When $B=0$, it will be driven by an harmonic field. When $A$ and $B$ $\neq0$ , it will be driven by a mixed field.
	
	The specific form of Hamiltonian $H_b$ is
	\begin{align}
		\label{eq_H}
		H_{b} = \begin{pmatrix}
			\Delta+g/4 & 0 & 0 & 0\\
			0 & -g/4  & g/2 & 0\\
			0 & g/2   & -g/4 & 0\\
			0 & 0 & 0 & g/4-\Delta\\
		\end{pmatrix},
	\end{align}
	Initially, each atom is in the ground state $|g\rangle$, and the whole system is in the lowest energy state,and the corresponding density matrix is $\rho(0)=|g\rangle \langle g|$. After passing through the drive of the field, the density matrix at the end time can be obtained\cite{PhysRevB.99.035421}
	\begin{align}
		\rho(t) = \begin{pmatrix}
			\frac{B^2}{4} & \frac{ - i B\sin 2A_1}{4} & \frac{ - i B \sin 2A_1}{4} & -\frac{AB}{4}\\
			\frac{-B\sin 2A_1}{4i} & \frac{\sin^2  2A_1}{4} & \frac{\sin^2  2A_1}{4} & \frac{A \sin 2A_1}{4i}\\
			\frac{-B\sin 2A_1}{4i} & \frac{\sin^2 2A_1}{4} & \frac{\sin^2 2A_1}{4} & \frac{A \sin 2A_1}{4i}\\
			-\frac{AB}{4} & \frac{i A \sin 2A_1}{4} & \frac{i A \sin 2A_1}{4} & \frac{A^2}{4}\\ 
		\end{pmatrix},
		\label{8}
	\end{align}
	where $ A=1+\cos 2A_1=1+r(t)$, $ B=1-\cos 2A_1=1-r(t)$, $ A_1=(Bt+(A/\omega)*\sin\omega t)/2$ and the final stored energy reads as
	\begin{align}
		E(t)= \Delta(1-r(t))
	\end{align}
	It is not difficult to see that when the interaction forces between atoms in all directions are consistent, the energy storage of the battery has no relationship with $g$. As shown in Figure~\ref{fig1}, the size of the battery energy storage is determined by the capacity $\Delta$ of the battery itself.
	
	In order to better observe the phenomenon of quantum steering and quantum entanglement, we take one of the two-level atoms as the observation object. After taking traces of the whole system, we obtained the reduced density matrix of the individual cells 
	\begin{align}
		\rho_b(t)=\begin{pmatrix}
			\frac{1-r(t)}{2}& 0 \\
			0 & \frac{1+r(t)}{2}
		\end{pmatrix},
		\label{5}
	\end{align}
	We can obtain the entangled entropy of the system by using Eq.(\ref{6}) and Eq.(\ref{5})
	\begin{align}
		S(t)=-\frac{1-r(t)}{2}\log_2\frac{1-r(t)}{2}-\frac{1+r(t)}{2}\log_2\frac{1+r(t)}{2},
	\end{align}
	And the ergotropy of a single battery cell can be obtained by Eq.(\ref{7}) and Eq.(\ref{5})
	
	\begin{align}
		\varepsilon(t)=\begin{cases}0,&r(t)\geq0\\-\Delta r(t),&r(t)<0
		\end{cases},
	\end{align}
	
	After calculation, we found that the energy storage of each battery cell is $E(t)/2$. and the ergotropy is $\varepsilon(t)$. Therefore, we can obtain the conversion efficiency of the whole system from a single battery cell.
	\begin{align}
		\eta(t)=\begin{cases}0,&r(t)\geq0\\2(1-\frac{1}{1-r(t)}),&r(t)<0
		\end{cases}
	\end{align}
	
	Then, we substitute Eq.(\ref{9}) into Eq.(\ref{10}) and Eq.(\ref{8}) to obtain the quantum-streeing $I^{ab}$ and $I^{ba}$ of the system. The specific parameters are as follows: $n_1=s_1=c_1=0$, $n_2=s_2=\sin2A_1$, $c_2=\sin^22A_1$, $n_3=-\cos2A_1$, $s_3=\sin^2A_1$, $c_3=\cos^22A_1$.
	
	In Figuer~\ref{fig1}, we observe that the hybrid field has a large increase in the charging rate of the QB compared to a single charging field. The minimum value of quantum entanglement corresponds to the maximum value of battery energy storage. In the steerable range, the maximum value of $a\rightarrow b$ quantum steering corresponds to the maximum value of battery energy storage and the minimum value of quantum entanglement. The $b\rightarrow a$ quantum steering minima corresponds to the maximum of battery energy storage and the minimum of quantum entanglement. This suggests that quantum steering may also be one of the indispensable resources for QB applications.
	
	Then, as we can see in Figuer~\ref{fig2}, the energy conversion rate of the quantum cell $\eta(t)\neq1$ for the harmonic field. The electrostatic field and the mixed field can make the conversion rate $\eta(t)=1$ at the moment when the energy storage reaches the maximum. In other words, the relationship between quantum entanglement and quantum steering and battery energy storage is equally applicable to the ergotropy of batteries.
	
	\begin{figure}[tb]
		\centering
		\includegraphics[width=1\linewidth]{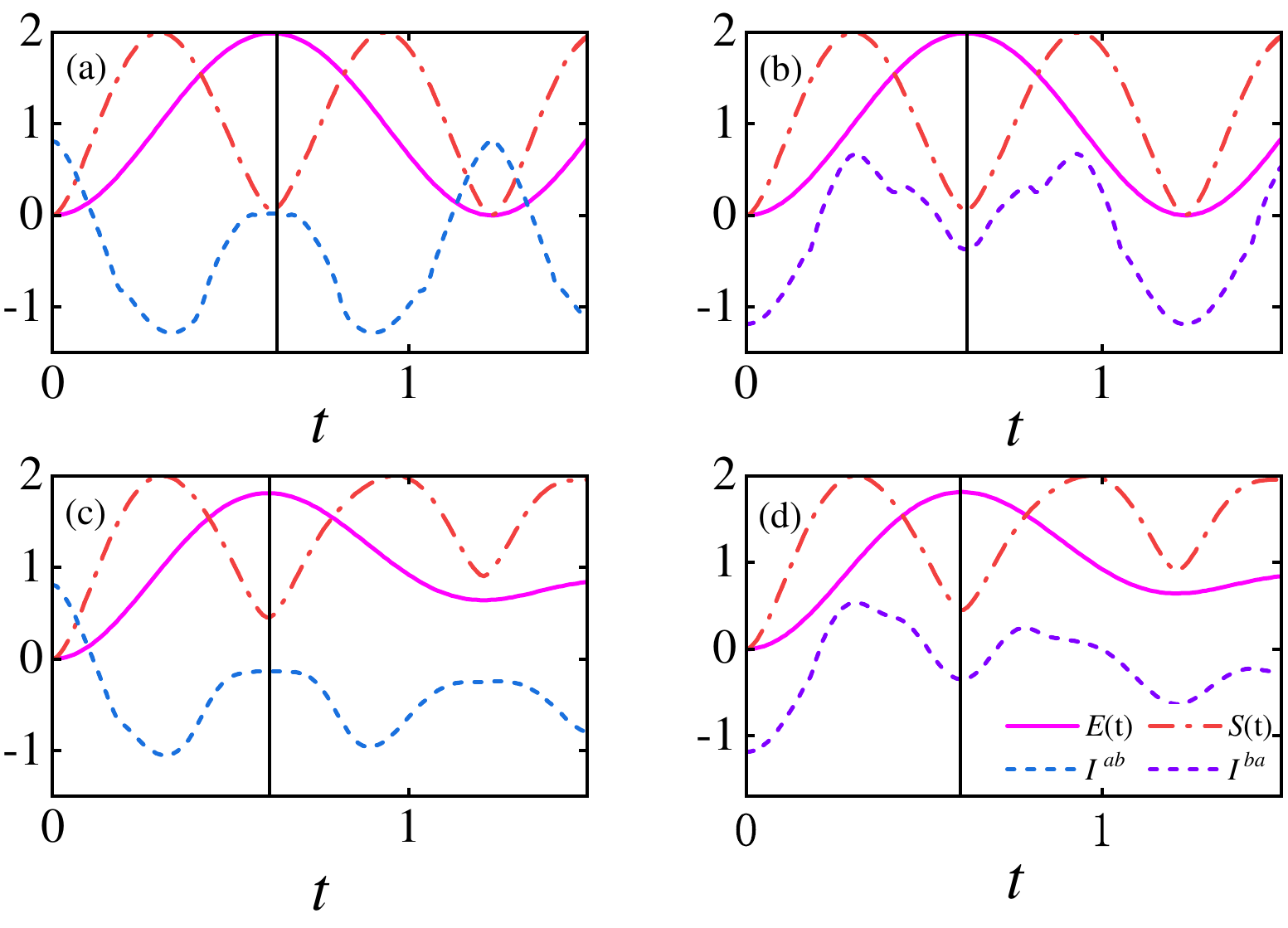}
		\caption{(a)-(d) plots the charger-battery entanglement $S(t)$, the $a\rightarrow b$ steering $I^{ab}$, $b\rightarrow a$ steering $I^{ba}$, and the stored energy $E(t)$ as a function of charging time $t$ for $n=7$, $\omega_{g}=1$,$J_1=1$, and $\Delta=1$. The anisotropy parameter $J_2=1$ in (a) and (b). The anisotropy parameter $J_2=0$ in (c) and (d). }
		\label{fig3}
	\end{figure}
	
	The reasons for this situation can be further analysed from Figure~\ref{fig5}. It shows that purity is a determinant of battery energy storage and ergotropy during charging. As the purity of the battery reaches the maximum, so does the corresponding energy storage and ergotropy. And when the purity is $1$, the conversion efficiency of the battery is $1$, and the ergotropy is at the maximum. This echoes the conclusion of \cite{shi2022entanglement}.
	\section{Cavity-Heisenberg chain QB}
	
	In order to make the results generalizable, we further analyzed the Cavity-Heisenberg chain QB\cite{dou2022cavity}. The Hamiltonian is 
	\begin{align}
		H_a=\omega_{a}a^{\dagger}a+\omega_{g}\sum_{i=1}^{2}\sigma _{i}^{x}(a^{\dagger}+a)
	\end{align}
	
	\begin{align}
		H_{b} =&\frac{\Delta }{2}\sum_{i=1}^{2}\sigma _{i}^{z}+%
		\sum_{i\neq j}^{2}(J_1\sigma _{i}^{x}\sigma _{j}^{x}+J_1\sigma _{i}^{y}\sigma
		_{j}^{y}+J_2\sigma _{i}^{z}\sigma
		_{j}^{z})  \notag \\
	\end{align}
	
	\begin{figure}[tb]
		\centering
		\includegraphics[width=0.85\linewidth]{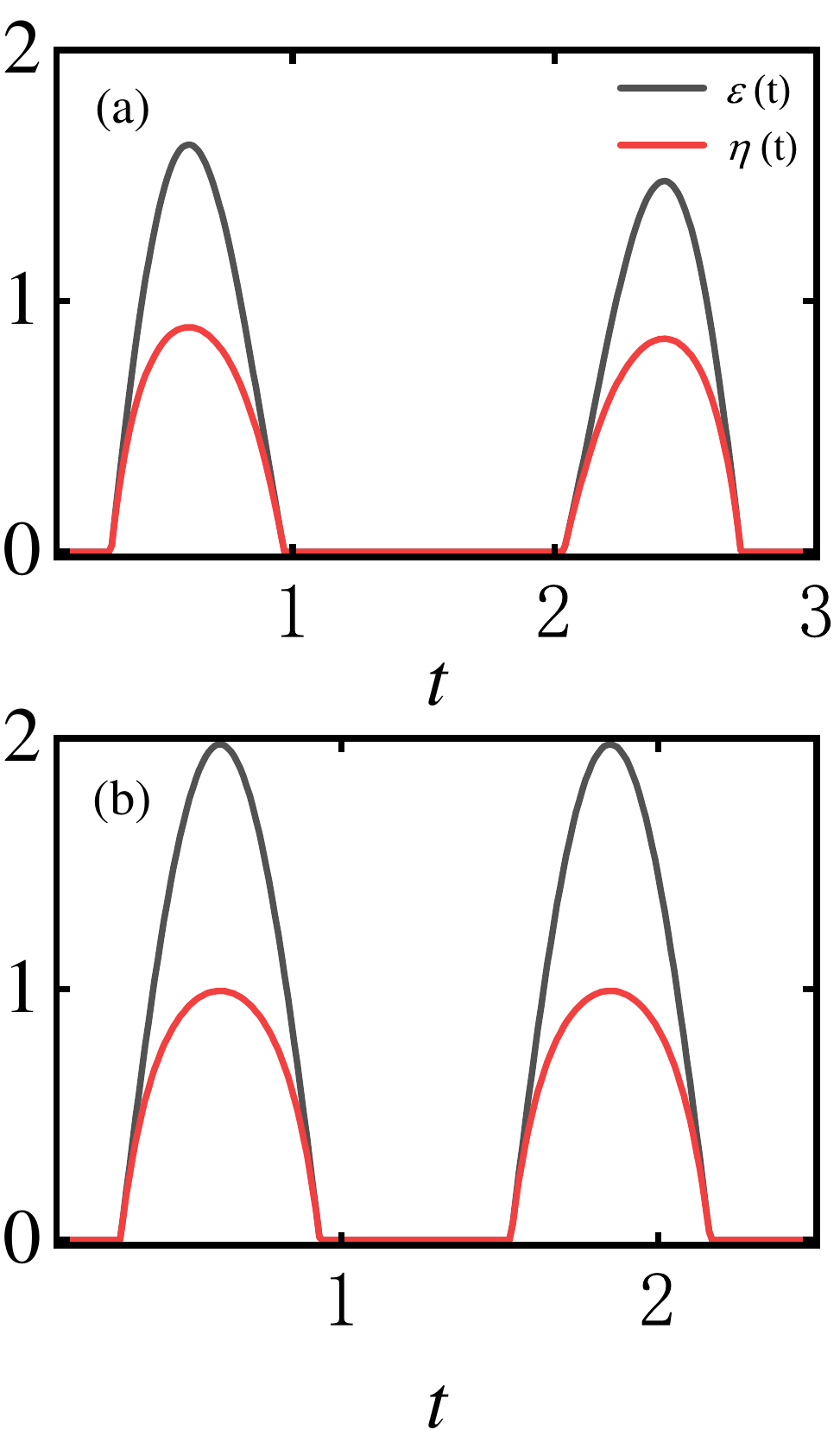}
		\caption{(a) and (b) plots the ergotropy $\varepsilon(t)$ and the efficiency of the conversion of work $\eta(t)$ as a function of charging time $t$ for $n=7$, $\omega_{g}=1$,$J_1=1$, and $\Delta=1$. The anisotropy parameters $J_2$ in (a) and (b) are $1$ and $0$, respectively. }
		\label{fig4}
	\end{figure}
	
	where $a$($a^{\dagger}$) are the annihilation (creation) operators of the cavity radiation, $\omega_a$ the frequency of the radiation in the cavity, and $\omega_{g}$ is the matter-radiation coupling. $J_1$ is the intensity of the interaction between atoms, $J_2$ is a dimensionless parameter
	associated with the anisotropy of the chain.
	We observe that the evolution of the system is determined by four states, namely: $|n,g,g\rangle$, $|n-1,e,g\rangle$, $|n-1,g,e\rangle$, $|n-2,e,e\rangle$. Correspondingly, the total Hamiltonian of the system can be expanded by them:
	\begin{align}
		H= \begin{pmatrix}
			n_1 & \omega_{g}\sqrt{n} &\omega_{g}\sqrt{n} &0 \\
			\omega_{g}\sqrt{n} & n_2 & 2J_1\Delta & \omega_{g}\sqrt{n-1}\\
			\omega_{g}\sqrt{n} & 2J_1\Delta &  n_3 & \omega_{g}\sqrt{n-1}\\
			0 & \omega_{g}\sqrt{n-1} & \omega_{g}\sqrt{n-1} & n_4\\ 
		\end{pmatrix},
	\end{align}
	where $n_1$=$n\omega_{a}+\Delta(J_2-1)$, $n_2$=$n_3$=$\omega_{a}(n-1)-J_2\Delta$, and $n_4$=$(n-2)\omega_{a}+\Delta(J_2+1)$.
	Here we We focus on the resonance
	regime, ie., $\omega_{a}=\Delta=1$, with the aim of maximizing the energy transfer in the charger-battery.
	
	Immediately after that, we plotted Figuer~\ref{fig3} and Figuer~\ref{fig4} according to Eq.(\ref{11}), Eq.(\ref{7}), Eq.(\ref{6}), Eq.(\ref{9}) and Eq.(\ref{10}).
	
	\begin{figure}[tb]
		\centering
		\includegraphics[width=1\linewidth]{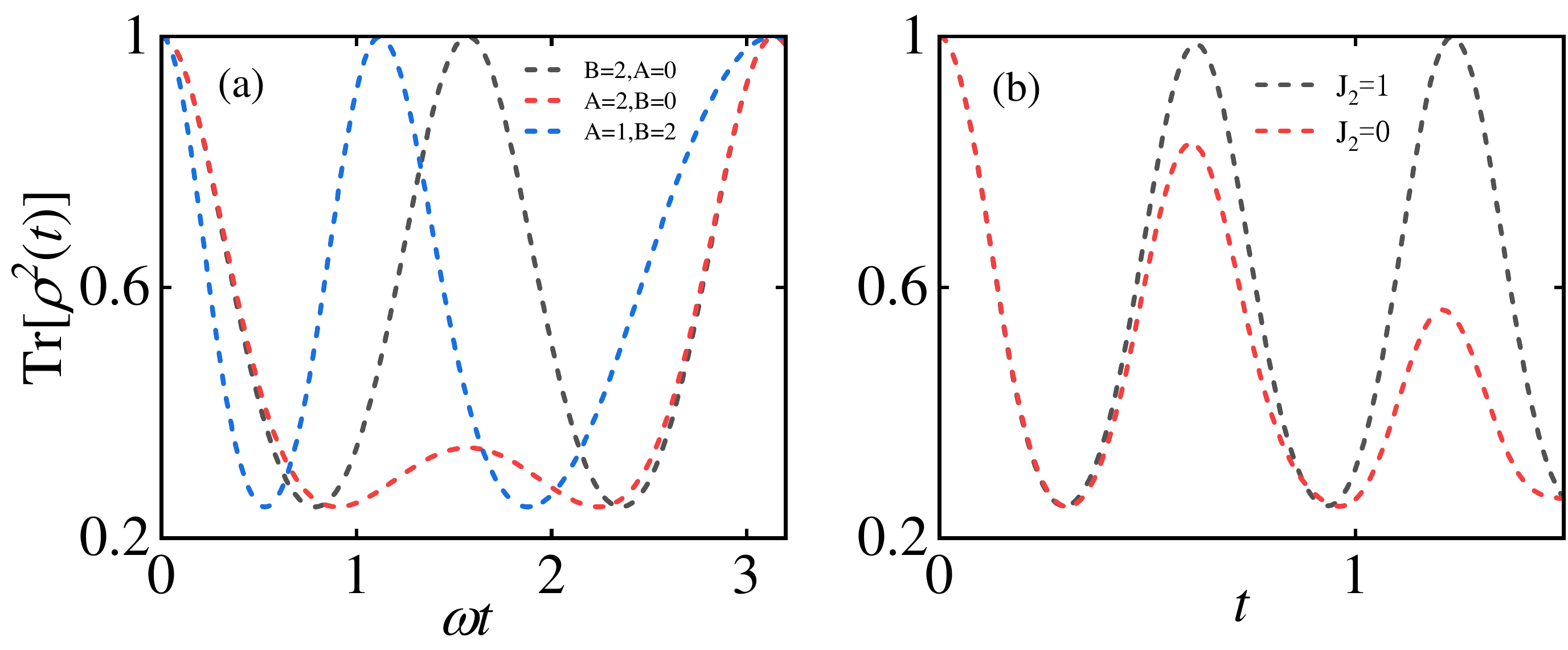}
		\caption{(a) Field-QB purity Tr$[\rho^2(t)]$ as a function of time $t$(in units of $\frac{1}{\omega}$). (b) Cavity-Heisenberg chain QB purity Tr$[\rho^2(t)]$ as a function of time $t$.  }
		\label{fig5}
	\end{figure}
	
	As shown in Figuer~\ref{fig3}, in the Cavity-Heisenberg chain model, the occurrence of the quantum entanglement minimum is always accompanied by the emergence of the maximum value of battery energy storage. In the steerable range, the $a\rightarrow b$ quantum steering is at the maximum, and the energy storage reaches the maximum. The $b\rightarrow a$ quantum steering is at the minimum, and energy storage is at the maximum. This result also shows that both quantum entanglement and quantum steering are indispensable resources in quantum batteries.
	
	By looking at Figuer~\ref{fig3} and Figuer~\ref{fig4}, we find that the energy storage of the battery reaches the maximum value, and the corresponding ergotropy is also at the maximum value. Therefore, the relationship between battery energy storage and quantum entanglement and quantum steering is also applicable to ergotropy. For the anisotropy of the spin chain, it will directly affect the ergotropy and energy storage of the QB. And when the anisotropic parameters $J_2=1$, the battery can reach a fully charged state. When $J_2\neq1$, the energy storage and ergotropy of the battery will decrease.
	
	In Figure~\ref{fig5}(b), we can also observe that when the purity of the system reaches a maximum, the energy storage of the corresponding battery is also maximum. At this time, the efficiency is the greatest, and the ergotropy is the largest.
	
	\section{conclusion}
	
	In this paper, we focus on the effects of quantum entanglement and entropy uncertainty-based quantum steering on Field-QB and Cavity-Heisenberg QB.  The relationship between system states and them is also discussed. We find that when quantum entanglement and quantum steering of $b\rightarrow a$ qubits are at a minimum, the energy storage and ergotropy of quantum batteries are always at a maximum. With the quantum steering of $a\rightarrow b$ qubits, the energy storage and ergotropy of batteries also appear maximums. This suggests that quantum steering is a resource that can be utilized in batteries. Immediately afterward, we analyzed the above reasons using the purity of the system. When the purity is maximum, the energy storage and ergotropy at this time are also maximum. When the purity is $1$, the energy storage and ergotropy are both maximal and equal, and the conversion rate is $1$. The reason is, the greater the purity of a quantum state, the smaller the entanglement in it. This is more generalizable than previous conclusions\cite{shi2022entanglement}. In addition to this, we also found that the anisotropy parameters in the spin chain are also key factors affecting battery energy storage. When the anisotropy parameter $J_2=1$, the battery can reach the state of complete charging. When $J_2\neq 1$, the battery does not reach that state. Our work add an available resource to the realization of QBs and deepen the understanding of quantum steering applications.
	
	\begin{acknowledgments}
		This work is supported by National Natural Science Foundations of China (No. 61975184), Science Foundation of Zhejiang Sci-Tech University (No. 19062151-Y).
	\end{acknowledgments}


\begin{thebibliography}{54}%
		\makeatletter
		\providecommand \@ifxundefined [1]{%
			\@ifx{#1\undefined}
		}%
		\providecommand \@ifnum [1]{%
			\ifnum #1\expandafter \@firstoftwo
			\else \expandafter \@secondoftwo
			\fi
		}%
		\providecommand \@ifx [1]{%
			\ifx #1\expandafter \@firstoftwo
			\else \expandafter \@secondoftwo
			\fi
		}%
		\providecommand \natexlab [1]{#1}%
		\providecommand \enquote  [1]{``#1''}%
		\providecommand \bibnamefont  [1]{#1}%
		\providecommand \bibfnamefont [1]{#1}%
		\providecommand \citenamefont [1]{#1}%
		\providecommand \href@noop [0]{\@secondoftwo}%
		\providecommand \href [0]{\begingroup \@sanitize@url \@href}%
		\providecommand \@href[1]{\@@startlink{#1}\@@href}%
		\providecommand \@@href[1]{\endgroup#1\@@endlink}%
		\providecommand \@sanitize@url [0]{\catcode `\\12\catcode `\$12\catcode
			`\&12\catcode `\#12\catcode `\^12\catcode `\_12\catcode `\%12\relax}%
		\providecommand \@@startlink[1]{}%
		\providecommand \@@endlink[0]{}%
		\providecommand \url  [0]{\begingroup\@sanitize@url \@url }%
		\providecommand \@url [1]{\endgroup\@href {#1}{\urlprefix }}%
		\providecommand \urlprefix  [0]{URL }%
		\providecommand \Eprint [0]{\href }%
		\providecommand \doibase [0]{https://doi.org/}%
		\providecommand \selectlanguage [0]{\@gobble}%
		\providecommand \bibinfo  [0]{\@secondoftwo}%
		\providecommand \bibfield  [0]{\@secondoftwo}%
		\providecommand \translation [1]{[#1]}%
		\providecommand \BibitemOpen [0]{}%
		\providecommand \bibitemStop [0]{}%
		\providecommand \bibitemNoStop [0]{.\EOS\space}%
		\providecommand \EOS [0]{\spacefactor3000\relax}%
		\providecommand \BibitemShut  [1]{\csname bibitem#1\endcsname}%
		\let\auto@bib@innerbib\@empty
		\bibitem [{\citenamefont {Pirandola}\ \emph {et~al.}(2018)\citenamefont {Pirandola}, \citenamefont {Bardhan}, \citenamefont {Gehring}, \citenamefont {Weedbrook},\ and\ \citenamefont {Lloyd}}]{pirandola2018advances}%
		\BibitemOpen
		\bibfield  {author} {\bibinfo {author} {\bibfnamefont {S.}~\bibnamefont {Pirandola}}, \bibinfo {author} {\bibfnamefont {B.~R.}\ \bibnamefont {Bardhan}}, \bibinfo {author} {\bibfnamefont {T.}~\bibnamefont {Gehring}}, \bibinfo {author} {\bibfnamefont {C.}~\bibnamefont {Weedbrook}},\ and\ \bibinfo {author} {\bibfnamefont {S.}~\bibnamefont {Lloyd}},\ }\href {https://doi.org/10.1038/s41566-018-0301-6} {\bibfield  {journal} {\bibinfo  {journal} {Nature Photonics}\ }\textbf {\bibinfo {volume} {12}},\ \bibinfo {pages} {724} (\bibinfo {year} {2018})}\BibitemShut {NoStop}%
		\bibitem [{\citenamefont {Cimini}\ \emph {et~al.}(2019)\citenamefont {Cimini}, \citenamefont {Gianani}, \citenamefont {Spagnolo}, \citenamefont {Leccese}, \citenamefont {Sciarrino},\ and\ \citenamefont {Barbieri}}]{PhysRevLett.123.230502}%
		\BibitemOpen
		\bibfield  {author} {\bibinfo {author} {\bibfnamefont {V.}~\bibnamefont {Cimini}}, \bibinfo {author} {\bibfnamefont {I.}~\bibnamefont {Gianani}}, \bibinfo {author} {\bibfnamefont {N.}~\bibnamefont {Spagnolo}}, \bibinfo {author} {\bibfnamefont {F.}~\bibnamefont {Leccese}}, \bibinfo {author} {\bibfnamefont {F.}~\bibnamefont {Sciarrino}},\ and\ \bibinfo {author} {\bibfnamefont {M.}~\bibnamefont {Barbieri}},\ }\href {https://doi.org/10.1103/PhysRevLett.123.230502} {\bibfield  {journal} {\bibinfo  {journal} {Phys. Rev. Lett.}\ }\textbf {\bibinfo {volume} {123}},\ \bibinfo {pages} {230502} (\bibinfo {year} {2019})}\BibitemShut {NoStop}%
		\bibitem [{\citenamefont {Quan}\ \emph {et~al.}(2007)\citenamefont {Quan}, \citenamefont {Liu}, \citenamefont {Sun},\ and\ \citenamefont {Nori}}]{PhysRevE.76.031105}%
		\BibitemOpen
		\bibfield  {author} {\bibinfo {author} {\bibfnamefont {H.~T.}\ \bibnamefont {Quan}}, \bibinfo {author} {\bibfnamefont {Y.-x.}\ \bibnamefont {Liu}}, \bibinfo {author} {\bibfnamefont {C.~P.}\ \bibnamefont {Sun}},\ and\ \bibinfo {author} {\bibfnamefont {F.}~\bibnamefont {Nori}},\ }\href {https://doi.org/10.1103/PhysRevE.76.031105} {\bibfield  {journal} {\bibinfo  {journal} {Phys. Rev. E}\ }\textbf {\bibinfo {volume} {76}},\ \bibinfo {pages} {031105} (\bibinfo {year} {2007})}\BibitemShut {NoStop}%
		\bibitem [{\citenamefont {Quan}(2009)}]{PhysRevE.79.041129}%
		\BibitemOpen
		\bibfield  {author} {\bibinfo {author} {\bibfnamefont {H.~T.}\ \bibnamefont {Quan}},\ }\href {https://doi.org/10.1103/PhysRevE.79.041129} {\bibfield  {journal} {\bibinfo  {journal} {Phys. Rev. E}\ }\textbf {\bibinfo {volume} {79}},\ \bibinfo {pages} {041129} (\bibinfo {year} {2009})}\BibitemShut {NoStop}%
		\bibitem [{\citenamefont {Peterson}\ \emph {et~al.}(2019)\citenamefont {Peterson}, \citenamefont {Batalh\~ao}, \citenamefont {Herrera}, \citenamefont {Souza}, \citenamefont {Sarthour}, \citenamefont {Oliveira},\ and\ \citenamefont {Serra}}]{PhysRevLett.123.240601}%
		\BibitemOpen
		\bibfield  {author} {\bibinfo {author} {\bibfnamefont {J.~P.~S.}\ \bibnamefont {Peterson}}, \bibinfo {author} {\bibfnamefont {T.~B.}\ \bibnamefont {Batalh\~ao}}, \bibinfo {author} {\bibfnamefont {M.}~\bibnamefont {Herrera}}, \bibinfo {author} {\bibfnamefont {A.~M.}\ \bibnamefont {Souza}}, \bibinfo {author} {\bibfnamefont {R.~S.}\ \bibnamefont {Sarthour}}, \bibinfo {author} {\bibfnamefont {I.~S.}\ \bibnamefont {Oliveira}},\ and\ \bibinfo {author} {\bibfnamefont {R.~M.}\ \bibnamefont {Serra}},\ }\href {https://doi.org/10.1103/PhysRevLett.123.240601} {\bibfield  {journal} {\bibinfo  {journal} {Phys. Rev. Lett.}\ }\textbf {\bibinfo {volume} {123}},\ \bibinfo {pages} {240601} (\bibinfo {year} {2019})}\BibitemShut {NoStop}%
		\bibitem [{\citenamefont {Denzler}\ and\ \citenamefont {Lutz}(2020)}]{PhysRevResearch.2.032062}%
		\BibitemOpen
		\bibfield  {author} {\bibinfo {author} {\bibfnamefont {T.}~\bibnamefont {Denzler}}\ and\ \bibinfo {author} {\bibfnamefont {E.}~\bibnamefont {Lutz}},\ }\href {https://doi.org/10.1103/PhysRevResearch.2.032062} {\bibfield  {journal} {\bibinfo  {journal} {Phys. Rev. Res.}\ }\textbf {\bibinfo {volume} {2}},\ \bibinfo {pages} {032062} (\bibinfo {year} {2020})}\BibitemShut {NoStop}%
		\bibitem [{\citenamefont {Degen}\ \emph {et~al.}(2017)\citenamefont {Degen}, \citenamefont {Reinhard},\ and\ \citenamefont {Cappellaro}}]{degen2017quantum}%
		\BibitemOpen
		\bibfield  {author} {\bibinfo {author} {\bibfnamefont {C.~L.}\ \bibnamefont {Degen}}, \bibinfo {author} {\bibfnamefont {F.}~\bibnamefont {Reinhard}},\ and\ \bibinfo {author} {\bibfnamefont {P.}~\bibnamefont {Cappellaro}},\ }\href {https://doi.org/10.1103/RevModPhys.89.035002} {\bibfield  {journal} {\bibinfo  {journal} {Rev. Mod. Phys.}\ }\textbf {\bibinfo {volume} {89}},\ \bibinfo {pages} {035002} (\bibinfo {year} {2017})}\BibitemShut {NoStop}%
		\bibitem [{\citenamefont {Marciniak}\ \emph {et~al.}(2022)\citenamefont {Marciniak}, \citenamefont {Feldker}, \citenamefont {Pogorelov}, \citenamefont {Kaubruegger}, \citenamefont {Vasilyev}, \citenamefont {van Bijnen}, \citenamefont {Schindler}, \citenamefont {Zoller}, \citenamefont {Blatt},\ and\ \citenamefont {Monz}}]{marciniak2022optimal}%
		\BibitemOpen
		\bibfield  {author} {\bibinfo {author} {\bibfnamefont {C.~D.}\ \bibnamefont {Marciniak}}, \bibinfo {author} {\bibfnamefont {T.}~\bibnamefont {Feldker}}, \bibinfo {author} {\bibfnamefont {I.}~\bibnamefont {Pogorelov}}, \bibinfo {author} {\bibfnamefont {R.}~\bibnamefont {Kaubruegger}}, \bibinfo {author} {\bibfnamefont {D.~V.}\ \bibnamefont {Vasilyev}}, \bibinfo {author} {\bibfnamefont {R.}~\bibnamefont {van Bijnen}}, \bibinfo {author} {\bibfnamefont {P.}~\bibnamefont {Schindler}}, \bibinfo {author} {\bibfnamefont {P.}~\bibnamefont {Zoller}}, \bibinfo {author} {\bibfnamefont {R.}~\bibnamefont {Blatt}},\ and\ \bibinfo {author} {\bibfnamefont {T.}~\bibnamefont {Monz}},\ }\href {https://doi.org/10.1038/s41586-022-04435-4} {\bibfield  {journal} {\bibinfo  {journal} {Nature}\ }\textbf {\bibinfo {volume} {603}},\ \bibinfo {pages} {604} (\bibinfo {year} {2022})}\BibitemShut {NoStop}%
		\bibitem [{\citenamefont {Boss}\ \emph {et~al.}(2017)\citenamefont {Boss}, \citenamefont {Cujia}, \citenamefont {Zopes},\ and\ \citenamefont {Degen}}]{boss2017quantum}%
		\BibitemOpen
		\bibfield  {author} {\bibinfo {author} {\bibfnamefont {J.~M.}\ \bibnamefont {Boss}}, \bibinfo {author} {\bibfnamefont {K.}~\bibnamefont {Cujia}}, \bibinfo {author} {\bibfnamefont {J.}~\bibnamefont {Zopes}},\ and\ \bibinfo {author} {\bibfnamefont {C.~L.}\ \bibnamefont {Degen}},\ }\href {DOI: 10.1126/science.aam7009} {\bibfield  {journal} {\bibinfo  {journal} {Science}\ }\textbf {\bibinfo {volume} {356}},\ \bibinfo {pages} {837} (\bibinfo {year} {2017})}\BibitemShut {NoStop}%
		\bibitem [{\citenamefont {Bass}\ and\ \citenamefont {Doser}(2024)}]{bass2024quantum}%
		\BibitemOpen
		\bibfield  {author} {\bibinfo {author} {\bibfnamefont {S.~D.}\ \bibnamefont {Bass}}\ and\ \bibinfo {author} {\bibfnamefont {M.}~\bibnamefont {Doser}},\ }\href {https://doi.org/10.1038/s42254-024-00714-3} {\bibfield  {journal} {\bibinfo  {journal} {Nature Reviews Physics}\ ,\ \bibinfo {pages} {1}} (\bibinfo {year} {2024})}\BibitemShut {NoStop}%
		\bibitem [{\citenamefont {Barra}(2019)}]{PhysRevLett.122.210601}%
		\BibitemOpen
		\bibfield  {author} {\bibinfo {author} {\bibfnamefont {F.}~\bibnamefont {Barra}},\ }\href {https://doi.org/10.1103/PhysRevLett.122.210601} {\bibfield  {journal} {\bibinfo  {journal} {Phys. Rev. Lett.}\ }\textbf {\bibinfo {volume} {122}},\ \bibinfo {pages} {210601} (\bibinfo {year} {2019})}\BibitemShut {NoStop}%
		\bibitem [{\citenamefont {Campaioli}\ \emph {et~al.}(2017)\citenamefont {Campaioli}, \citenamefont {Pollock}, \citenamefont {Binder}, \citenamefont {C\'eleri}, \citenamefont {Goold}, \citenamefont {Vinjanampathy},\ and\ \citenamefont {Modi}}]{PhysRevLett.118.150601}%
		\BibitemOpen
		\bibfield  {author} {\bibinfo {author} {\bibfnamefont {F.}~\bibnamefont {Campaioli}}, \bibinfo {author} {\bibfnamefont {F.~A.}\ \bibnamefont {Pollock}}, \bibinfo {author} {\bibfnamefont {F.~C.}\ \bibnamefont {Binder}}, \bibinfo {author} {\bibfnamefont {L.}~\bibnamefont {C\'eleri}}, \bibinfo {author} {\bibfnamefont {J.}~\bibnamefont {Goold}}, \bibinfo {author} {\bibfnamefont {S.}~\bibnamefont {Vinjanampathy}},\ and\ \bibinfo {author} {\bibfnamefont {K.}~\bibnamefont {Modi}},\ }\href {https://doi.org/10.1103/PhysRevLett.118.150601} {\bibfield  {journal} {\bibinfo  {journal} {Phys. Rev. Lett.}\ }\textbf {\bibinfo {volume} {118}},\ \bibinfo {pages} {150601} (\bibinfo {year} {2017})}\BibitemShut {NoStop}%
		\bibitem [{\citenamefont {Jiang}\ \emph {et~al.}(2022)\citenamefont {Jiang}, \citenamefont {Chen}, \citenamefont {Xiao}, \citenamefont {Pan}, \citenamefont {Jin}, \citenamefont {Yu},\ and\ \citenamefont {Chen}}]{jiang2022quantum}%
		\BibitemOpen
		\bibfield  {author} {\bibinfo {author} {\bibfnamefont {Y.}~\bibnamefont {Jiang}}, \bibinfo {author} {\bibfnamefont {T.}~\bibnamefont {Chen}}, \bibinfo {author} {\bibfnamefont {C.}~\bibnamefont {Xiao}}, \bibinfo {author} {\bibfnamefont {K.}~\bibnamefont {Pan}}, \bibinfo {author} {\bibfnamefont {G.}~\bibnamefont {Jin}}, \bibinfo {author} {\bibfnamefont {Y.}~\bibnamefont {Yu}},\ and\ \bibinfo {author} {\bibfnamefont {A.}~\bibnamefont {Chen}},\ }\href {https://doi.org/10.3390/e24121821} {\bibfield  {journal} {\bibinfo  {journal} {Entropy}\ }\textbf {\bibinfo {volume} {24}},\ \bibinfo {pages} {1821} (\bibinfo {year} {2022})}\BibitemShut {NoStop}%
		\bibitem [{\citenamefont {Alicki}\ and\ \citenamefont {Fannes}(2013)}]{alicki2013entanglement}%
		\BibitemOpen
		\bibfield  {author} {\bibinfo {author} {\bibfnamefont {R.}~\bibnamefont {Alicki}}\ and\ \bibinfo {author} {\bibfnamefont {M.}~\bibnamefont {Fannes}},\ }\href {https://doi.org/10.1103/PhysRevE.87.042123} {\bibfield  {journal} {\bibinfo  {journal} {Phys. Rev. E}\ }\textbf {\bibinfo {volume} {87}},\ \bibinfo {pages} {042123} (\bibinfo {year} {2013})}\BibitemShut {NoStop}%
		\bibitem [{\citenamefont {Carrasco}\ \emph {et~al.}(2022)\citenamefont {Carrasco}, \citenamefont {Maze}, \citenamefont {Hermann-Avigliano},\ and\ \citenamefont {Barra}}]{carrasco2110collective}%
		\BibitemOpen
		\bibfield  {author} {\bibinfo {author} {\bibfnamefont {J.}~\bibnamefont {Carrasco}}, \bibinfo {author} {\bibfnamefont {J.~R.}\ \bibnamefont {Maze}}, \bibinfo {author} {\bibfnamefont {C.}~\bibnamefont {Hermann-Avigliano}},\ and\ \bibinfo {author} {\bibfnamefont {F.}~\bibnamefont {Barra}},\ }\href {https://doi.org/10.1103/PhysRevE.105.064119} {\bibfield  {journal} {\bibinfo  {journal} {Phys. Rev. E}\ }\textbf {\bibinfo {volume} {105}},\ \bibinfo {pages} {064119} (\bibinfo {year} {2022})}\BibitemShut {NoStop}%
		\bibitem [{\citenamefont {Dou}\ \emph {et~al.}(2022)\citenamefont {Dou}, \citenamefont {Zhou},\ and\ \citenamefont {Sun}}]{dou2022cavity}%
		\BibitemOpen
		\bibfield  {author} {\bibinfo {author} {\bibfnamefont {F.-Q.}\ \bibnamefont {Dou}}, \bibinfo {author} {\bibfnamefont {H.}~\bibnamefont {Zhou}},\ and\ \bibinfo {author} {\bibfnamefont {J.-A.}\ \bibnamefont {Sun}},\ }\href {https://doi.org/10.1103/PhysRevA.106.032212} {\bibfield  {journal} {\bibinfo  {journal} {Physical Review A}\ }\textbf {\bibinfo {volume} {106}},\ \bibinfo {pages} {032212} (\bibinfo {year} {2022})}\BibitemShut {NoStop}%
		\bibitem [{\citenamefont {Zhang}\ \emph {et~al.}(2019)\citenamefont {Zhang}, \citenamefont {Yang}, \citenamefont {Fu},\ and\ \citenamefont {Wang}}]{zhang2019powerful}%
		\BibitemOpen
		\bibfield  {author} {\bibinfo {author} {\bibfnamefont {Y.-Y.}\ \bibnamefont {Zhang}}, \bibinfo {author} {\bibfnamefont {T.-R.}\ \bibnamefont {Yang}}, \bibinfo {author} {\bibfnamefont {L.}~\bibnamefont {Fu}},\ and\ \bibinfo {author} {\bibfnamefont {X.}~\bibnamefont {Wang}},\ }\href {https://doi.org/10.1103/PhysRevE.99.052106} {\bibfield  {journal} {\bibinfo  {journal} {Phys. Rev. E}\ }\textbf {\bibinfo {volume} {99}},\ \bibinfo {pages} {052106} (\bibinfo {year} {2019})}\BibitemShut {NoStop}%
		\bibitem [{\citenamefont {Yang}\ \emph {et~al.}(2023)\citenamefont {Yang}, \citenamefont {Yang}, \citenamefont {Alimuddin}, \citenamefont {Salvia}, \citenamefont {Fei}, \citenamefont {Zhao}, \citenamefont {Nimmrichter},\ and\ \citenamefont {Luo}}]{PhysRevLett.131.030402}%
		\BibitemOpen
		\bibfield  {author} {\bibinfo {author} {\bibfnamefont {X.}~\bibnamefont {Yang}}, \bibinfo {author} {\bibfnamefont {Y.-H.}\ \bibnamefont {Yang}}, \bibinfo {author} {\bibfnamefont {M.}~\bibnamefont {Alimuddin}}, \bibinfo {author} {\bibfnamefont {R.}~\bibnamefont {Salvia}}, \bibinfo {author} {\bibfnamefont {S.-M.}\ \bibnamefont {Fei}}, \bibinfo {author} {\bibfnamefont {L.-M.}\ \bibnamefont {Zhao}}, \bibinfo {author} {\bibfnamefont {S.}~\bibnamefont {Nimmrichter}},\ and\ \bibinfo {author} {\bibfnamefont {M.-X.}\ \bibnamefont {Luo}},\ }\href {https://doi.org/10.1103/PhysRevLett.131.030402} {\bibfield  {journal} {\bibinfo  {journal} {Phys. Rev. Lett.}\ }\textbf {\bibinfo {volume} {131}},\ \bibinfo {pages} {030402} (\bibinfo {year} {2023})}\BibitemShut {NoStop}%
		\bibitem [{\citenamefont {Pirmoradian}\ and\ \citenamefont {M\o{}lmer}(2019)}]{pirmoradian2019aging}%
		\BibitemOpen
		\bibfield  {author} {\bibinfo {author} {\bibfnamefont {F.}~\bibnamefont {Pirmoradian}}\ and\ \bibinfo {author} {\bibfnamefont {K.}~\bibnamefont {M\o{}lmer}},\ }\href {https://doi.org/10.1103/PhysRevA.100.043833} {\bibfield  {journal} {\bibinfo  {journal} {Phys. Rev. A}\ }\textbf {\bibinfo {volume} {100}},\ \bibinfo {pages} {043833} (\bibinfo {year} {2019})}\BibitemShut {NoStop}%
		\bibitem [{\citenamefont {Carrega}\ \emph {et~al.}(2020)\citenamefont {Carrega}, \citenamefont {Crescente}, \citenamefont {Ferraro},\ and\ \citenamefont {Sassetti}}]{carrega2020dissipative}%
		\BibitemOpen
		\bibfield  {author} {\bibinfo {author} {\bibfnamefont {M.}~\bibnamefont {Carrega}}, \bibinfo {author} {\bibfnamefont {A.}~\bibnamefont {Crescente}}, \bibinfo {author} {\bibfnamefont {D.}~\bibnamefont {Ferraro}},\ and\ \bibinfo {author} {\bibfnamefont {M.}~\bibnamefont {Sassetti}},\ }\href {https://doi.org/10.1088/1367-2630/abaa01} {\bibfield  {journal} {\bibinfo  {journal} {New Journal of Physics}\ }\textbf {\bibinfo {volume} {22}},\ \bibinfo {pages} {083085} (\bibinfo {year} {2020})}\BibitemShut {NoStop}%
		\bibitem [{\citenamefont {Qu}\ \emph {et~al.}(2023)\citenamefont {Qu}, \citenamefont {Zhan}, \citenamefont {Lin},\ and\ \citenamefont {Xue}}]{qu2023experimental}%
		\BibitemOpen
		\bibfield  {author} {\bibinfo {author} {\bibfnamefont {D.}~\bibnamefont {Qu}}, \bibinfo {author} {\bibfnamefont {X.}~\bibnamefont {Zhan}}, \bibinfo {author} {\bibfnamefont {H.}~\bibnamefont {Lin}},\ and\ \bibinfo {author} {\bibfnamefont {P.}~\bibnamefont {Xue}},\ }\href {https://doi.org/10.1103/PhysRevB.108.L180301} {\bibfield  {journal} {\bibinfo  {journal} {Physical Review B}\ }\textbf {\bibinfo {volume} {108}},\ \bibinfo {pages} {L180301} (\bibinfo {year} {2023})}\BibitemShut {NoStop}%
		\bibitem [{\citenamefont {Joshi}\ and\ \citenamefont {Mahesh}(2022)}]{joshi2022experimental}%
		\BibitemOpen
		\bibfield  {author} {\bibinfo {author} {\bibfnamefont {J.}~\bibnamefont {Joshi}}\ and\ \bibinfo {author} {\bibfnamefont {T.}~\bibnamefont {Mahesh}},\ }\href {https://doi.org/10.1103/PhysRevA.106.042601} {\bibfield  {journal} {\bibinfo  {journal} {Physical Review A}\ }\textbf {\bibinfo {volume} {106}},\ \bibinfo {pages} {042601} (\bibinfo {year} {2022})}\BibitemShut {NoStop}%
		\bibitem [{\citenamefont {Le}\ \emph {et~al.}(2018)\citenamefont {Le}, \citenamefont {Levinsen}, \citenamefont {Modi}, \citenamefont {Parish},\ and\ \citenamefont {Pollock}}]{le2018spin}%
		\BibitemOpen
		\bibfield  {author} {\bibinfo {author} {\bibfnamefont {T.~P.}\ \bibnamefont {Le}}, \bibinfo {author} {\bibfnamefont {J.}~\bibnamefont {Levinsen}}, \bibinfo {author} {\bibfnamefont {K.}~\bibnamefont {Modi}}, \bibinfo {author} {\bibfnamefont {M.~M.}\ \bibnamefont {Parish}},\ and\ \bibinfo {author} {\bibfnamefont {F.~A.}\ \bibnamefont {Pollock}},\ }\href {https://doi.org/10.1103/PhysRevA.97.022106} {\bibfield  {journal} {\bibinfo  {journal} {Phys. Rev. A}\ }\textbf {\bibinfo {volume} {97}},\ \bibinfo {pages} {022106} (\bibinfo {year} {2018})}\BibitemShut {NoStop}%
		\bibitem [{\citenamefont {Ferraro}\ \emph {et~al.}(2018)\citenamefont {Ferraro}, \citenamefont {Campisi}, \citenamefont {Andolina}, \citenamefont {Pellegrini},\ and\ \citenamefont {Polini}}]{ferraro2018high}%
		\BibitemOpen
		\bibfield  {author} {\bibinfo {author} {\bibfnamefont {D.}~\bibnamefont {Ferraro}}, \bibinfo {author} {\bibfnamefont {M.}~\bibnamefont {Campisi}}, \bibinfo {author} {\bibfnamefont {G.~M.}\ \bibnamefont {Andolina}}, \bibinfo {author} {\bibfnamefont {V.}~\bibnamefont {Pellegrini}},\ and\ \bibinfo {author} {\bibfnamefont {M.}~\bibnamefont {Polini}},\ }\href {https://doi.org/10.1103/PhysRevLett.120.117702} {\bibfield  {journal} {\bibinfo  {journal} {Phys. Rev. Lett.}\ }\textbf {\bibinfo {volume} {120}},\ \bibinfo {pages} {117702} (\bibinfo {year} {2018})}\BibitemShut {NoStop}%
		\bibitem [{\citenamefont {Kamin}\ \emph {et~al.}(2020)\citenamefont {Kamin}, \citenamefont {Tabesh}, \citenamefont {Salimi},\ and\ \citenamefont {Santos}}]{PhysRevE.102.052109}%
		\BibitemOpen
		\bibfield  {author} {\bibinfo {author} {\bibfnamefont {F.~H.}\ \bibnamefont {Kamin}}, \bibinfo {author} {\bibfnamefont {F.~T.}\ \bibnamefont {Tabesh}}, \bibinfo {author} {\bibfnamefont {S.}~\bibnamefont {Salimi}},\ and\ \bibinfo {author} {\bibfnamefont {A.~C.}\ \bibnamefont {Santos}},\ }\href {https://doi.org/10.1103/PhysRevE.102.052109} {\bibfield  {journal} {\bibinfo  {journal} {Phys. Rev. E}\ }\textbf {\bibinfo {volume} {102}},\ \bibinfo {pages} {052109} (\bibinfo {year} {2020})}\BibitemShut {NoStop}%
		\bibitem [{\citenamefont {Liu}\ \emph {et~al.}(2021)\citenamefont {Liu}, \citenamefont {Shi}, \citenamefont {Shi}, \citenamefont {Wang},\ and\ \citenamefont {Yang}}]{PhysRevB.104.245418}%
		\BibitemOpen
		\bibfield  {author} {\bibinfo {author} {\bibfnamefont {J.-X.}\ \bibnamefont {Liu}}, \bibinfo {author} {\bibfnamefont {H.-L.}\ \bibnamefont {Shi}}, \bibinfo {author} {\bibfnamefont {Y.-H.}\ \bibnamefont {Shi}}, \bibinfo {author} {\bibfnamefont {X.-H.}\ \bibnamefont {Wang}},\ and\ \bibinfo {author} {\bibfnamefont {W.-L.}\ \bibnamefont {Yang}},\ }\href {https://doi.org/10.1103/PhysRevB.104.245418} {\bibfield  {journal} {\bibinfo  {journal} {Phys. Rev. B}\ }\textbf {\bibinfo {volume} {104}},\ \bibinfo {pages} {245418} (\bibinfo {year} {2021})}\BibitemShut {NoStop}%
		\bibitem [{\citenamefont {Andolina}\ \emph {et~al.}(2019{\natexlab{a}})\citenamefont {Andolina}, \citenamefont {Keck}, \citenamefont {Mari}, \citenamefont {Campisi}, \citenamefont {Giovannetti},\ and\ \citenamefont {Polini}}]{andolina2019extractable}%
		\BibitemOpen
		\bibfield  {author} {\bibinfo {author} {\bibfnamefont {G.~M.}\ \bibnamefont {Andolina}}, \bibinfo {author} {\bibfnamefont {M.}~\bibnamefont {Keck}}, \bibinfo {author} {\bibfnamefont {A.}~\bibnamefont {Mari}}, \bibinfo {author} {\bibfnamefont {M.}~\bibnamefont {Campisi}}, \bibinfo {author} {\bibfnamefont {V.}~\bibnamefont {Giovannetti}},\ and\ \bibinfo {author} {\bibfnamefont {M.}~\bibnamefont {Polini}},\ }\href {https://doi.org/10.1103/PhysRevLett.122.047702} {\bibfield  {journal} {\bibinfo  {journal} {Phys. Rev. Lett.}\ }\textbf {\bibinfo {volume} {122}},\ \bibinfo {pages} {047702} (\bibinfo {year} {2019}{\natexlab{a}})}\BibitemShut {NoStop}%
		\bibitem [{\citenamefont {Shi}\ \emph {et~al.}(2022)\citenamefont {Shi}, \citenamefont {Ding}, \citenamefont {Wan}, \citenamefont {Wang},\ and\ \citenamefont {Yang}}]{shi2022entanglement}%
		\BibitemOpen
		\bibfield  {author} {\bibinfo {author} {\bibfnamefont {H.-L.}\ \bibnamefont {Shi}}, \bibinfo {author} {\bibfnamefont {S.}~\bibnamefont {Ding}}, \bibinfo {author} {\bibfnamefont {Q.-K.}\ \bibnamefont {Wan}}, \bibinfo {author} {\bibfnamefont {X.-H.}\ \bibnamefont {Wang}},\ and\ \bibinfo {author} {\bibfnamefont {W.-L.}\ \bibnamefont {Yang}},\ }\href {https://doi.org/10.1103/PhysRevLett.129.130602} {\bibfield  {journal} {\bibinfo  {journal} {Phys. Rev. Lett.}\ }\textbf {\bibinfo {volume} {129}},\ \bibinfo {pages} {130602} (\bibinfo {year} {2022})}\BibitemShut {NoStop}%
		\bibitem [{\citenamefont {Yao}\ and\ \citenamefont {Shao}(2021)}]{PhysRevE.104.044116}%
		\BibitemOpen
		\bibfield  {author} {\bibinfo {author} {\bibfnamefont {Y.}~\bibnamefont {Yao}}\ and\ \bibinfo {author} {\bibfnamefont {X.~Q.}\ \bibnamefont {Shao}},\ }\href {https://doi.org/10.1103/PhysRevE.104.044116} {\bibfield  {journal} {\bibinfo  {journal} {Phys. Rev. E}\ }\textbf {\bibinfo {volume} {104}},\ \bibinfo {pages} {044116} (\bibinfo {year} {2021})}\BibitemShut {NoStop}%
		\bibitem [{\citenamefont {Kamin}\ \emph {et~al.}(2024)\citenamefont {Kamin}, \citenamefont {Salimi},\ and\ \citenamefont {Arjmandi}}]{PhysRevA.109.022226}%
		\BibitemOpen
		\bibfield  {author} {\bibinfo {author} {\bibfnamefont {F.~H.}\ \bibnamefont {Kamin}}, \bibinfo {author} {\bibfnamefont {S.}~\bibnamefont {Salimi}},\ and\ \bibinfo {author} {\bibfnamefont {M.~B.}\ \bibnamefont {Arjmandi}},\ }\href {https://doi.org/10.1103/PhysRevA.109.022226} {\bibfield  {journal} {\bibinfo  {journal} {Phys. Rev. A}\ }\textbf {\bibinfo {volume} {109}},\ \bibinfo {pages} {022226} (\bibinfo {year} {2024})}\BibitemShut {NoStop}%
		\bibitem [{\citenamefont {Crescente}\ \emph {et~al.}(2020)\citenamefont {Crescente}, \citenamefont {Carrega}, \citenamefont {Sassetti},\ and\ \citenamefont {Ferraro}}]{PhysRevB.102.245407}%
		\BibitemOpen
		\bibfield  {author} {\bibinfo {author} {\bibfnamefont {A.}~\bibnamefont {Crescente}}, \bibinfo {author} {\bibfnamefont {M.}~\bibnamefont {Carrega}}, \bibinfo {author} {\bibfnamefont {M.}~\bibnamefont {Sassetti}},\ and\ \bibinfo {author} {\bibfnamefont {D.}~\bibnamefont {Ferraro}},\ }\href {https://doi.org/10.1103/PhysRevB.102.245407} {\bibfield  {journal} {\bibinfo  {journal} {Phys. Rev. B}\ }\textbf {\bibinfo {volume} {102}},\ \bibinfo {pages} {245407} (\bibinfo {year} {2020})}\BibitemShut {NoStop}%
		\bibitem [{\citenamefont {Dicke}(1954)}]{dicke1954coherence}%
		\BibitemOpen
		\bibfield  {author} {\bibinfo {author} {\bibfnamefont {R.~H.}\ \bibnamefont {Dicke}},\ }\href {https://doi.org/10.1103/PhysRev.93.99} {\bibfield  {journal} {\bibinfo  {journal} {Phys. Rev.}\ }\textbf {\bibinfo {volume} {93}},\ \bibinfo {pages} {99} (\bibinfo {year} {1954})}\BibitemShut {NoStop}%
		\bibitem [{\citenamefont {Baumann}\ \emph {et~al.}(2010)\citenamefont {Baumann}, \citenamefont {Guerlin}, \citenamefont {Brennecke},\ and\ \citenamefont {Esslinger}}]{baumann2010dicke}%
		\BibitemOpen
		\bibfield  {author} {\bibinfo {author} {\bibfnamefont {K.}~\bibnamefont {Baumann}}, \bibinfo {author} {\bibfnamefont {C.}~\bibnamefont {Guerlin}}, \bibinfo {author} {\bibfnamefont {F.}~\bibnamefont {Brennecke}},\ and\ \bibinfo {author} {\bibfnamefont {T.}~\bibnamefont {Esslinger}},\ }\href {https://doi.org/10.1038/nature09009} {\bibfield  {journal} {\bibinfo  {journal} {nature}\ }\textbf {\bibinfo {volume} {464}},\ \bibinfo {pages} {1301} (\bibinfo {year} {2010})}\BibitemShut {NoStop}%
		\bibitem [{\citenamefont {Zhang}\ \emph {et~al.}(2023)\citenamefont {Zhang}, \citenamefont {Wang}, \citenamefont {Wu},\ and\ \citenamefont {Wang}}]{PhysRevE.107.054125}%
		\BibitemOpen
		\bibfield  {author} {\bibinfo {author} {\bibfnamefont {W.}~\bibnamefont {Zhang}}, \bibinfo {author} {\bibfnamefont {S.}~\bibnamefont {Wang}}, \bibinfo {author} {\bibfnamefont {C.}~\bibnamefont {Wu}},\ and\ \bibinfo {author} {\bibfnamefont {G.}~\bibnamefont {Wang}},\ }\href {https://doi.org/10.1103/PhysRevE.107.054125} {\bibfield  {journal} {\bibinfo  {journal} {Phys. Rev. E}\ }\textbf {\bibinfo {volume} {107}},\ \bibinfo {pages} {054125} (\bibinfo {year} {2023})}\BibitemShut {NoStop}%
		\bibitem [{\citenamefont {Horodecki}\ \emph {et~al.}(2009)\citenamefont {Horodecki}, \citenamefont {Horodecki}, \citenamefont {Horodecki},\ and\ \citenamefont {Horodecki}}]{RevModPhys.81.865}%
		\BibitemOpen
		\bibfield  {author} {\bibinfo {author} {\bibfnamefont {R.}~\bibnamefont {Horodecki}}, \bibinfo {author} {\bibfnamefont {P.}~\bibnamefont {Horodecki}}, \bibinfo {author} {\bibfnamefont {M.}~\bibnamefont {Horodecki}},\ and\ \bibinfo {author} {\bibfnamefont {K.}~\bibnamefont {Horodecki}},\ }\href {https://doi.org/10.1103/RevModPhys.81.865} {\bibfield  {journal} {\bibinfo  {journal} {Rev. Mod. Phys.}\ }\textbf {\bibinfo {volume} {81}},\ \bibinfo {pages} {865} (\bibinfo {year} {2009})}\BibitemShut {NoStop}%
		\bibitem [{\citenamefont {Streltsov}\ \emph {et~al.}(2017)\citenamefont {Streltsov}, \citenamefont {Adesso},\ and\ \citenamefont {Plenio}}]{RevModPhys.89.041003}%
		\BibitemOpen
		\bibfield  {author} {\bibinfo {author} {\bibfnamefont {A.}~\bibnamefont {Streltsov}}, \bibinfo {author} {\bibfnamefont {G.}~\bibnamefont {Adesso}},\ and\ \bibinfo {author} {\bibfnamefont {M.~B.}\ \bibnamefont {Plenio}},\ }\href {https://doi.org/10.1103/RevModPhys.89.041003} {\bibfield  {journal} {\bibinfo  {journal} {Rev. Mod. Phys.}\ }\textbf {\bibinfo {volume} {89}},\ \bibinfo {pages} {041003} (\bibinfo {year} {2017})}\BibitemShut {NoStop}%
		\bibitem [{\citenamefont {Andolina}\ \emph {et~al.}(2019{\natexlab{b}})\citenamefont {Andolina}, \citenamefont {Keck}, \citenamefont {Mari}, \citenamefont {Campisi}, \citenamefont {Giovannetti},\ and\ \citenamefont {Polini}}]{PhysRevLett.122.047702}%
		\BibitemOpen
		\bibfield  {author} {\bibinfo {author} {\bibfnamefont {G.~M.}\ \bibnamefont {Andolina}}, \bibinfo {author} {\bibfnamefont {M.}~\bibnamefont {Keck}}, \bibinfo {author} {\bibfnamefont {A.}~\bibnamefont {Mari}}, \bibinfo {author} {\bibfnamefont {M.}~\bibnamefont {Campisi}}, \bibinfo {author} {\bibfnamefont {V.}~\bibnamefont {Giovannetti}},\ and\ \bibinfo {author} {\bibfnamefont {M.}~\bibnamefont {Polini}},\ }\href {https://doi.org/10.1103/PhysRevLett.122.047702} {\bibfield  {journal} {\bibinfo  {journal} {Phys. Rev. Lett.}\ }\textbf {\bibinfo {volume} {122}},\ \bibinfo {pages} {047702} (\bibinfo {year} {2019}{\natexlab{b}})}\BibitemShut {NoStop}%
		\bibitem [{\citenamefont {Garc\'{\i}a-Pintos}\ \emph {et~al.}(2020)\citenamefont {Garc\'{\i}a-Pintos}, \citenamefont {Hamma},\ and\ \citenamefont {del Campo}}]{PhysRevLett.125.040601}%
		\BibitemOpen
		\bibfield  {author} {\bibinfo {author} {\bibfnamefont {L.~P.}\ \bibnamefont {Garc\'{\i}a-Pintos}}, \bibinfo {author} {\bibfnamefont {A.}~\bibnamefont {Hamma}},\ and\ \bibinfo {author} {\bibfnamefont {A.}~\bibnamefont {del Campo}},\ }\href {https://doi.org/10.1103/PhysRevLett.125.040601} {\bibfield  {journal} {\bibinfo  {journal} {Phys. Rev. Lett.}\ }\textbf {\bibinfo {volume} {125}},\ \bibinfo {pages} {040601} (\bibinfo {year} {2020})}\BibitemShut {NoStop}%
		\bibitem [{\citenamefont {Liu}\ \emph {et~al.}(2020)\citenamefont {Liu}, \citenamefont {Liang}, \citenamefont {Jin}, \citenamefont {Yu}, \citenamefont {Lan}, \citenamefont {He},\ and\ \citenamefont {Guo}}]{Liu_2020}%
		\BibitemOpen
		\bibfield  {author} {\bibinfo {author} {\bibfnamefont {Y.}~\bibnamefont {Liu}}, \bibinfo {author} {\bibfnamefont {S.-L.}\ \bibnamefont {Liang}}, \bibinfo {author} {\bibfnamefont {G.-R.}\ \bibnamefont {Jin}}, \bibinfo {author} {\bibfnamefont {Y.-B.}\ \bibnamefont {Yu}}, \bibinfo {author} {\bibfnamefont {J.-Y.}\ \bibnamefont {Lan}}, \bibinfo {author} {\bibfnamefont {X.-B.}\ \bibnamefont {He}},\ and\ \bibinfo {author} {\bibfnamefont {K.-X.}\ \bibnamefont {Guo}},\ }\href {https://doi.org/10.1088/1674-1056/ab7da6} {\bibfield  {journal} {\bibinfo  {journal} {Chinese Physics B}\ }\textbf {\bibinfo {volume} {29}},\ \bibinfo {pages} {050301} (\bibinfo {year} {2020})}\BibitemShut {NoStop}%
		\bibitem [{\citenamefont {Uola}\ \emph {et~al.}(2020)\citenamefont {Uola}, \citenamefont {Costa}, \citenamefont {Nguyen},\ and\ \citenamefont {G\"uhne}}]{RevModPhys.92.015001}%
		\BibitemOpen
		\bibfield  {author} {\bibinfo {author} {\bibfnamefont {R.}~\bibnamefont {Uola}}, \bibinfo {author} {\bibfnamefont {A.~C.~S.}\ \bibnamefont {Costa}}, \bibinfo {author} {\bibfnamefont {H.~C.}\ \bibnamefont {Nguyen}},\ and\ \bibinfo {author} {\bibfnamefont {O.}~\bibnamefont {G\"uhne}},\ }\href {https://doi.org/10.1103/RevModPhys.92.015001} {\bibfield  {journal} {\bibinfo  {journal} {Rev. Mod. Phys.}\ }\textbf {\bibinfo {volume} {92}},\ \bibinfo {pages} {015001} (\bibinfo {year} {2020})}\BibitemShut {NoStop}%
		\bibitem [{\citenamefont {Kogias}\ \emph {et~al.}(2015)\citenamefont {Kogias}, \citenamefont {Lee}, \citenamefont {Ragy},\ and\ \citenamefont {Adesso}}]{PhysRevLett.114.060403}%
		\BibitemOpen
		\bibfield  {author} {\bibinfo {author} {\bibfnamefont {I.}~\bibnamefont {Kogias}}, \bibinfo {author} {\bibfnamefont {A.~R.}\ \bibnamefont {Lee}}, \bibinfo {author} {\bibfnamefont {S.}~\bibnamefont {Ragy}},\ and\ \bibinfo {author} {\bibfnamefont {G.}~\bibnamefont {Adesso}},\ }\href {https://doi.org/10.1103/PhysRevLett.114.060403} {\bibfield  {journal} {\bibinfo  {journal} {Phys. Rev. Lett.}\ }\textbf {\bibinfo {volume} {114}},\ \bibinfo {pages} {060403} (\bibinfo {year} {2015})}\BibitemShut {NoStop}%
		\bibitem [{\citenamefont {Fan}\ \emph {et~al.}(2022)\citenamefont {Fan}, \citenamefont {Jia},\ and\ \citenamefont {Qiu}}]{PhysRevA.106.012433}%
		\BibitemOpen
		\bibfield  {author} {\bibinfo {author} {\bibfnamefont {Y.}~\bibnamefont {Fan}}, \bibinfo {author} {\bibfnamefont {C.}~\bibnamefont {Jia}},\ and\ \bibinfo {author} {\bibfnamefont {L.}~\bibnamefont {Qiu}},\ }\href {https://doi.org/10.1103/PhysRevA.106.012433} {\bibfield  {journal} {\bibinfo  {journal} {Phys. Rev. A}\ }\textbf {\bibinfo {volume} {106}},\ \bibinfo {pages} {012433} (\bibinfo {year} {2022})}\BibitemShut {NoStop}%
		\bibitem [{\citenamefont {Guo}\ \emph {et~al.}(2019)\citenamefont {Guo}, \citenamefont {Cheng}, \citenamefont {Hu}, \citenamefont {Liu}, \citenamefont {Huang}, \citenamefont {Huang}, \citenamefont {Li}, \citenamefont {Guo},\ and\ \citenamefont {Cavalcanti}}]{PhysRevLett.123.170402}%
		\BibitemOpen
		\bibfield  {author} {\bibinfo {author} {\bibfnamefont {Y.}~\bibnamefont {Guo}}, \bibinfo {author} {\bibfnamefont {S.}~\bibnamefont {Cheng}}, \bibinfo {author} {\bibfnamefont {X.}~\bibnamefont {Hu}}, \bibinfo {author} {\bibfnamefont {B.-H.}\ \bibnamefont {Liu}}, \bibinfo {author} {\bibfnamefont {E.-M.}\ \bibnamefont {Huang}}, \bibinfo {author} {\bibfnamefont {Y.-F.}\ \bibnamefont {Huang}}, \bibinfo {author} {\bibfnamefont {C.-F.}\ \bibnamefont {Li}}, \bibinfo {author} {\bibfnamefont {G.-C.}\ \bibnamefont {Guo}},\ and\ \bibinfo {author} {\bibfnamefont {E.~G.}\ \bibnamefont {Cavalcanti}},\ }\href {https://doi.org/10.1103/PhysRevLett.123.170402} {\bibfield  {journal} {\bibinfo  {journal} {Phys. Rev. Lett.}\ }\textbf {\bibinfo {volume} {123}},\ \bibinfo {pages} {170402} (\bibinfo {year} {2019})}\BibitemShut {NoStop}%
		\bibitem [{\citenamefont {Sun}\ \emph {et~al.}(2017)\citenamefont {Sun}, \citenamefont {Wang}, \citenamefont {Shi},\ and\ \citenamefont {Ye}}]{sun2017exploration}%
		\BibitemOpen
		\bibfield  {author} {\bibinfo {author} {\bibfnamefont {W.-Y.}\ \bibnamefont {Sun}}, \bibinfo {author} {\bibfnamefont {D.}~\bibnamefont {Wang}}, \bibinfo {author} {\bibfnamefont {J.-D.}\ \bibnamefont {Shi}},\ and\ \bibinfo {author} {\bibfnamefont {L.}~\bibnamefont {Ye}},\ }\href {https://doi.org/10.1038/srep39651} {\bibfield  {journal} {\bibinfo  {journal} {Scientific Reports}\ }\textbf {\bibinfo {volume} {7}},\ \bibinfo {pages} {39651} (\bibinfo {year} {2017})}\BibitemShut {NoStop}%
		\bibitem [{\citenamefont {Reid}(2013)}]{PhysRevA.88.062338}%
		\BibitemOpen
		\bibfield  {author} {\bibinfo {author} {\bibfnamefont {M.~D.}\ \bibnamefont {Reid}},\ }\href {https://doi.org/10.1103/PhysRevA.88.062338} {\bibfield  {journal} {\bibinfo  {journal} {Phys. Rev. A}\ }\textbf {\bibinfo {volume} {88}},\ \bibinfo {pages} {062338} (\bibinfo {year} {2013})}\BibitemShut {NoStop}%
		\bibitem [{\citenamefont {Nagy}\ and\ \citenamefont {V{\'e}rtesi}(2016)}]{nagy2016epr}%
		\BibitemOpen
		\bibfield  {author} {\bibinfo {author} {\bibfnamefont {S.}~\bibnamefont {Nagy}}\ and\ \bibinfo {author} {\bibfnamefont {T.}~\bibnamefont {V{\'e}rtesi}},\ }\href {https://doi.org/https://doi.org/10.1038/srep21634} {\bibfield  {journal} {\bibinfo  {journal} {Scientific reports}\ }\textbf {\bibinfo {volume} {6}},\ \bibinfo {pages} {21634} (\bibinfo {year} {2016})}\BibitemShut {NoStop}%
		\bibitem [{\citenamefont {Chiu}\ \emph {et~al.}(2016)\citenamefont {Chiu}, \citenamefont {Lambert}, \citenamefont {Liao}, \citenamefont {Nori},\ and\ \citenamefont {Li}}]{chiu2016no}%
		\BibitemOpen
		\bibfield  {author} {\bibinfo {author} {\bibfnamefont {C.-Y.}\ \bibnamefont {Chiu}}, \bibinfo {author} {\bibfnamefont {N.}~\bibnamefont {Lambert}}, \bibinfo {author} {\bibfnamefont {T.-L.}\ \bibnamefont {Liao}}, \bibinfo {author} {\bibfnamefont {F.}~\bibnamefont {Nori}},\ and\ \bibinfo {author} {\bibfnamefont {C.-M.}\ \bibnamefont {Li}},\ }\href {https://doi.org/https://doi.org/10.1038/npjqi.2016.20} {\bibfield  {journal} {\bibinfo  {journal} {NPJ Quantum Information}\ }\textbf {\bibinfo {volume} {2}},\ \bibinfo {pages} {1} (\bibinfo {year} {2016})}\BibitemShut {NoStop}%
		\bibitem [{\citenamefont {Schneeloch}\ \emph {et~al.}(2013{\natexlab{a}})\citenamefont {Schneeloch}, \citenamefont {Broadbent}, \citenamefont {Walborn}, \citenamefont {Cavalcanti},\ and\ \citenamefont {Howell}}]{PhysRevA.87.062103}%
		\BibitemOpen
		\bibfield  {author} {\bibinfo {author} {\bibfnamefont {J.}~\bibnamefont {Schneeloch}}, \bibinfo {author} {\bibfnamefont {C.~J.}\ \bibnamefont {Broadbent}}, \bibinfo {author} {\bibfnamefont {S.~P.}\ \bibnamefont {Walborn}}, \bibinfo {author} {\bibfnamefont {E.~G.}\ \bibnamefont {Cavalcanti}},\ and\ \bibinfo {author} {\bibfnamefont {J.~C.}\ \bibnamefont {Howell}},\ }\href {https://doi.org/10.1103/PhysRevA.87.062103} {\bibfield  {journal} {\bibinfo  {journal} {Phys. Rev. A}\ }\textbf {\bibinfo {volume} {87}},\ \bibinfo {pages} {062103} (\bibinfo {year} {2013}{\natexlab{a}})}\BibitemShut {NoStop}%
		\bibitem [{\citenamefont {Farina}\ \emph {et~al.}(2019)\citenamefont {Farina}, \citenamefont {Andolina}, \citenamefont {Mari}, \citenamefont {Polini},\ and\ \citenamefont {Giovannetti}}]{PhysRevB.99.035421}%
		\BibitemOpen
		\bibfield  {author} {\bibinfo {author} {\bibfnamefont {D.}~\bibnamefont {Farina}}, \bibinfo {author} {\bibfnamefont {G.~M.}\ \bibnamefont {Andolina}}, \bibinfo {author} {\bibfnamefont {A.}~\bibnamefont {Mari}}, \bibinfo {author} {\bibfnamefont {M.}~\bibnamefont {Polini}},\ and\ \bibinfo {author} {\bibfnamefont {V.}~\bibnamefont {Giovannetti}},\ }\href {https://doi.org/10.1103/PhysRevB.99.035421} {\bibfield  {journal} {\bibinfo  {journal} {Phys. Rev. B}\ }\textbf {\bibinfo {volume} {99}},\ \bibinfo {pages} {035421} (\bibinfo {year} {2019})}\BibitemShut {NoStop}%
		\bibitem [{\citenamefont {Ghosh}\ \emph {et~al.}(2020)\citenamefont {Ghosh}, \citenamefont {Chanda},\ and\ \citenamefont {Sen(De)}}]{PhysRevA.101.032115}%
		\BibitemOpen
		\bibfield  {author} {\bibinfo {author} {\bibfnamefont {S.}~\bibnamefont {Ghosh}}, \bibinfo {author} {\bibfnamefont {T.}~\bibnamefont {Chanda}},\ and\ \bibinfo {author} {\bibfnamefont {A.}~\bibnamefont {Sen(De)}},\ }\href {https://doi.org/10.1103/PhysRevA.101.032115} {\bibfield  {journal} {\bibinfo  {journal} {Phys. Rev. A}\ }\textbf {\bibinfo {volume} {101}},\ \bibinfo {pages} {032115} (\bibinfo {year} {2020})}\BibitemShut {NoStop}%
		\bibitem [{\citenamefont {Van~Horne}\ \emph {et~al.}(2020)\citenamefont {Van~Horne}, \citenamefont {Yum}, \citenamefont {Dutta}, \citenamefont {H{\"a}nggi}, \citenamefont {Gong}, \citenamefont {Poletti},\ and\ \citenamefont {Mukherjee}}]{van2020single}%
		\BibitemOpen
		\bibfield  {author} {\bibinfo {author} {\bibfnamefont {N.}~\bibnamefont {Van~Horne}}, \bibinfo {author} {\bibfnamefont {D.}~\bibnamefont {Yum}}, \bibinfo {author} {\bibfnamefont {T.}~\bibnamefont {Dutta}}, \bibinfo {author} {\bibfnamefont {P.}~\bibnamefont {H{\"a}nggi}}, \bibinfo {author} {\bibfnamefont {J.}~\bibnamefont {Gong}}, \bibinfo {author} {\bibfnamefont {D.}~\bibnamefont {Poletti}},\ and\ \bibinfo {author} {\bibfnamefont {M.}~\bibnamefont {Mukherjee}},\ }\href {https://doi.org/https://doi.org/10.1038/s41534-020-0264-6} {\bibfield  {journal} {\bibinfo  {journal} {npj Quantum Information}\ }\textbf {\bibinfo {volume} {6}},\ \bibinfo {pages} {37} (\bibinfo {year} {2020})}\BibitemShut {NoStop}%
		\bibitem [{\citenamefont {von Lindenfels}\ \emph {et~al.}(2019)\citenamefont {von Lindenfels}, \citenamefont {Gr\"ab}, \citenamefont {Schmiegelow}, \citenamefont {Kaushal}, \citenamefont {Schulz}, \citenamefont {Mitchison}, \citenamefont {Goold}, \citenamefont {Schmidt-Kaler},\ and\ \citenamefont {Poschinger}}]{PhysRevLett.123.080602}%
		\BibitemOpen
		\bibfield  {author} {\bibinfo {author} {\bibfnamefont {D.}~\bibnamefont {von Lindenfels}}, \bibinfo {author} {\bibfnamefont {O.}~\bibnamefont {Gr\"ab}}, \bibinfo {author} {\bibfnamefont {C.~T.}\ \bibnamefont {Schmiegelow}}, \bibinfo {author} {\bibfnamefont {V.}~\bibnamefont {Kaushal}}, \bibinfo {author} {\bibfnamefont {J.}~\bibnamefont {Schulz}}, \bibinfo {author} {\bibfnamefont {M.~T.}\ \bibnamefont {Mitchison}}, \bibinfo {author} {\bibfnamefont {J.}~\bibnamefont {Goold}}, \bibinfo {author} {\bibfnamefont {F.}~\bibnamefont {Schmidt-Kaler}},\ and\ \bibinfo {author} {\bibfnamefont {U.~G.}\ \bibnamefont {Poschinger}},\ }\href {https://doi.org/10.1103/PhysRevLett.123.080602} {\bibfield  {journal} {\bibinfo  {journal} {Phys. Rev. Lett.}\ }\textbf {\bibinfo {volume} {123}},\ \bibinfo {pages} {080602} (\bibinfo {year} {2019})}\BibitemShut {NoStop}%
		\bibitem [{\citenamefont {Vedral}\ \emph {et~al.}(1997)\citenamefont {Vedral}, \citenamefont {Plenio}, \citenamefont {Rippin},\ and\ \citenamefont {Knight}}]{PhysRevLett.78.2275}%
		\BibitemOpen
		\bibfield  {author} {\bibinfo {author} {\bibfnamefont {V.}~\bibnamefont {Vedral}}, \bibinfo {author} {\bibfnamefont {M.~B.}\ \bibnamefont {Plenio}}, \bibinfo {author} {\bibfnamefont {M.~A.}\ \bibnamefont {Rippin}},\ and\ \bibinfo {author} {\bibfnamefont {P.~L.}\ \bibnamefont {Knight}},\ }\href {https://doi.org/10.1103/PhysRevLett.78.2275} {\bibfield  {journal} {\bibinfo  {journal} {Phys. Rev. Lett.}\ }\textbf {\bibinfo {volume} {78}},\ \bibinfo {pages} {2275} (\bibinfo {year} {1997})}\BibitemShut {NoStop}%
		\bibitem [{\citenamefont {Schneeloch}\ \emph {et~al.}(2013{\natexlab{b}})\citenamefont {Schneeloch}, \citenamefont {Dixon}, \citenamefont {Howland}, \citenamefont {Broadbent},\ and\ \citenamefont {Howell}}]{PhysRevLett.110.130407}%
		\BibitemOpen
		\bibfield  {author} {\bibinfo {author} {\bibfnamefont {J.}~\bibnamefont {Schneeloch}}, \bibinfo {author} {\bibfnamefont {P.~B.}\ \bibnamefont {Dixon}}, \bibinfo {author} {\bibfnamefont {G.~A.}\ \bibnamefont {Howland}}, \bibinfo {author} {\bibfnamefont {C.~J.}\ \bibnamefont {Broadbent}},\ and\ \bibinfo {author} {\bibfnamefont {J.~C.}\ \bibnamefont {Howell}},\ }\href {https://doi.org/10.1103/PhysRevLett.110.130407} {\bibfield  {journal} {\bibinfo  {journal} {Phys. Rev. Lett.}\ }\textbf {\bibinfo {volume} {110}},\ \bibinfo {pages} {130407} (\bibinfo {year} {2013}{\natexlab{b}})}\BibitemShut {NoStop}%
	\end{thebibliography}
\end{document}